\documentclass[prd,amsmath,amssymb,showpacs,superscriptaddress,nofootinbib,twocolumn]{revtex4-1}
\RequirePackage[T1]{fontenc}

\RequirePackage{graphicx}
\RequirePackage{mathptmx}      % use Times fonts if available on your TeX system

\usepackage{graphicx}
\usepackage{epsfig,graphics,subfigure,psfrag,amsmath,amssymb}
\usepackage{lineno}
\usepackage{dcolumn}
\usepackage{bm}
\usepackage{overpic}
\usepackage{xspace}
\usepackage{float}

\usepackage{color}
\usepackage{multirow}
\usepackage{colortbl}
\usepackage{epstopdf}
\usepackage[colorlinks,linkcolor=blue,anchorcolor=blue,citecolor=blue]{hyperref}
\usepackage{dcolumn}
\usepackage{amsmath}

\newcolumntype{C}{D{\pm}{\pm}{3, 5}}

\makeatletter
\renewcommand*\env@matrix[1][*\c@MaxMatrixCols c]{%
  \hskip -\arraycolsep
  \let\@ifnextchar\new@ifnextchar
  \array{#1}}
\makeatother

\definecolor{arsenic}{rgb}{1.0, 0.34, 0.27}

\begin{document}

\title{Data-taking strategy for the precise measurement of the $W$ boson mass with threshold scan at circular electron positron colliders}
%\title{A template for CEPC papers}
%\draftversion{1.0}

%==================================== Authors ===================================================
\author{P.~X.~Shen}
\thanks{Electronic address: shenpx@mail.nankai.edu.cn}
\affiliation{Nankai University, Tianjin 300071, People's Republic of China}
\author{P. Azzurri}
\affiliation{INFN, sezione di Pisa, Italy}
\author{C.~X.~Yu}
\thanks{Electronic address: yucx@nankai.edu.cn}
\affiliation{Nankai University, Tianjin 300071, People's Republic of China}
\author{M.~Boonekamp}
\affiliation{IRFU, CEA, Universite Paris-Saclay, Paris}
\author{C. M.~Kuo}
\affiliation{Department of Physics and Center for High Energy and High Field Physics, National Central University, Tao yuan}
\author{P.~Z.~Lai}
\affiliation{Department of Physics and Center for High Energy and High Field Physics, National Central University, Tao yuan}
\author{B.~Li}
\affiliation{Department of Physics, Yantai University, Yantai, Shandong}
\author{G.~Li}
\thanks{Electronic address: li.gang@mail.ihep.ac.cn}
\affiliation{Institute of High Energy Physics, Beijing 100049, People's Republic of China}
\author{H.~N.~Li}
\affiliation{South China Normal University, Guangzhou, Guangdong}
\author{Z.~J.~Liang}
\affiliation{Institute of High Energy Physics, Beijing 100049, People's Republic of China}
\author{B.~Liu}
\affiliation{Institute of High Energy Physics, Beijing 100049, People's Republic of China}
\author{J.~M.~Qian}
\affiliation{Department of Physics, University of Michigan, Ann Arbor, MI}
\author{L.~S.~Shi}
\affiliation{School of Physics, Sun Yat-sen University, Guangzhou, Guangdong}

\begin{abstract}
Circular electron positron colliders, such as the CEPC and FCC-ee, have been proposed
to measure Higgs boson properties precisely, test the Standard Model, search for physics
beyond the Standard Model, and so on. One of the important goals of these colliders is to measure
the $W$ boson mass with great precision by taking data around the $W$-pair production threshold. In this paper,
the data-taking scheme is investigated to maximize the achievable precisions of the $W$ boson mass and width
with a threshold scan, when various systematic uncertainties are taken into account.
The study shows that an optimal and realistic data-taking scheme is to collect data
at three center-of-mass energies and that precisions of 1.0~MeV and 3.4~MeV can be achieved for the mass
and width of the $W$ boson, respectively, with a total integrated luminosity of $\mathcal{L}=3.2$~\mbox{ab}$^{-1}$
and several assumptions of the systematic uncertainty sources.
\end{abstract}

\maketitle

%%%%%%%%%%%%%%%%%%%%%%%%%%%%%%%%%%%%%%%%%%%%%%%%%%%%%%%%%%%%%%%%%%%%%%%%%%%%%%%
% This is where the document really begins
%%%%%%%%%%%%%%%%%%%%%%%%%%%%%%%%%%%%%%%%%%%%%%%%%%%%%%%%%%%%%%%%%%%%%%%%%%%%%%%

% Shorthand for \phantom to use in tables
\newcommand{\pho}{\phantom{0}}
\newcommand{\bslash}{\ensuremath{\backslash}}
\newcommand{\BibTeX}{{\sc Bib\TeX}}

%% ============================================== Introduction =============================================
\section{Introduction}
In the Standard Model (SM) of particle physics, the electroweak (EW) interaction is mediated by the
$W$ boson, the $Z$ boson, and the photon, in a gauge theory based on the $\rm{SU(2)_{L}}\times\rm{U(1)_{Y}}$
symmetry~\cite{Gauge_1,Gauge_2,Gauge_3}. The so called symmetry-breaking
mechanism is based on the interaction of the gauge bosons with a scalar doublet field and predicts
the existence of a new physical state known as the Higgs boson~\cite{Higgs_1,Higgs_2,Higgs_3}.
The $W$ and $Z$ bosons were discovered by the UA1 and UA2 Collaborations in
1983~\cite{WZ_boson_1,WZ_boson_2,WZ_boson_3,WZ_boson_4} and the Higgs boson was discovered by the
ATLAS and CMS Collaborations in 2012~\cite{Higgs_E_1,Higgs_E_2}.

In the EW theory, the $W$ boson mass, $m_{W}$, can be expressed as a function of the $Z$ boson mass, $m_{Z}$;
the fine-structure constant, $\alpha$; the Fermi constant, $G_{\mu}$; the top-quark mass, $m_{t}$; and the Higgs boson mass, $m_{H}$. With the measured values of these parameters, the SM predicted value of the $W$ boson mass has been calculated to be $80.358\pm0.008$ GeV in Ref.~\cite{WMass_1} and $80.362\pm0.008$ GeV in Ref.~\cite{WMass_2}. The current Particle Data Group (PDG) world average value of $m_{W}=80.379\pm0.012$ MeV~\cite{WMass_PDG2018} is dominated by the measurements at LEP2 and Tevatron as well as the latest measurement by the ATLAS Collaboration.
In the context of global fits to the SM parameters, constraints on physics beyond the SM are currently limited by
the precision of $m_{W}$, $m_{t}$, and $m_{H}$. High precision measurements of these masses are essential to test
the overall consistency of the SM and search for new physics beyond the SM.

There are several methods to measure the $m_{W}$, as proposed for the LEP2 program~\cite{WMass_TH_LEP1,WMass_TH_LEP2,HSCheng,Stirling_1,Stirling_2}.
The first one is the direct reconstruction method, with kinematically-constrained or mass reconstructions of $W^+W^-$,
which is the most used in the current experimental results of both hadron and lepton colliders.
This method suffers from large systematic uncertainties such as those from
hadronization modeling, radiative corrections, lepton energy scale, missing energy, and so on.
The second method for $m_{W}$ measuring is to use the lepton end-point energy. This method encounters the lepton energy calibration problem.
The third method is that the $W$ boson mass can be determined by comparing the observed
$W$-pair production cross section(s) ($\sigma_{WW}$) near their kinematic threshold.
%with the theoretical prediction(s) of the EW theory. This is because $\sigma_{WW}$ near the $W$-pair threshold is very sensitive to the mass and width ($\Gamma_W$) of the $W$ boson.
Just like the measurement of the $\tau$ lepton mass~\cite{Tau_1,Tau_2}, this method has potential to measure the $m_{W}$ with
high precision when collecting large data sample around the $W^+W^-$ threshold. Based on this strategy,
LEP2 experiments have measured the $W$-pair cross section at a single energy point near 161~GeV,
with a total integrated luminosity of about 10~$\mbox{pb}^{-1}$ for each
of the four experiments. The $W$ boson mass was determined with a precision of
200~MeV~\cite{ALEPH,DELPHI,L3,OPAL}, dominated by the statistical uncertainty.
With much larger data samples, the precision of the $W$ boson mass using this method is expected to be improved significantly.
%The advantage of this method is that it is only sensitive to the number of events, so the $W$ boson mass can be determined with a high precision from a large data sample around the $W$-pair production threshold.
After the discovery of the Higgs boson~\cite{Higgs_1,Higgs_2,Higgs_3}, several large electron positron colliders have been proposed,
such as the ILC~\cite{ILC}, FCC-ee~\cite{Fcc_ee,EPOL17} and CEPC~\cite{CEPC_1}. One of their important physics goals is the precise measurement
of the $W$ boson mass. With the expected high integrated luminosity, the threshold scan method is well suited.

For the linear colliders, the $W$-pair threshold scan using polarized beams have been studied by TESLA physics program~\cite{TESLA},
and ILC~\cite{ILC_WMass}. The different polarization states have a advantage to enhance  signal cross section and to measure the background in situ.
For the circular colliders, the concept of a multi-point scan of the W threshold to extract mass and width of the W boson,
and the related data-taking optimization strategy was introduced in the context of FCC-ee
studies~\cite{Fcc_ee_WMass}, which reveal that an optimal strategy
would include measuring $\sigma_{WW}$ at the $\Gamma_W$-independent energy point $\sqrt{s}\simeq 2m_W + 1.5$~GeV,
and ``off-shell'' at $\sqrt{s}\simeq 2m_W$ - (1--2)$\Gamma_W$.
Scenarios where systematic uncertainties would be limiting the precision have been examined separately
for different sources, and provided the indication that systematic effects that are correlated at different
energy points could be partially canceled by measuring $\sigma_{WW}$ at additional energy
points where the differential coefficients relevant to the systematic uncertainties
are equal~\cite{FCCW16Azzurri,FCCW17Azzurri,FCCW18Azzurri,EPOL17Janot,EPOL17Azzurri}.

In this paper we follow the same methodology, extending it to the context of the CEPC planned data-taking,
and produce comparisons with current FCC-ee projections.
Additional care and insight is given to a comprehensive evaluation of the impact and possible limitations
of systematic uncertainties on the final measurements.

The threshold scan method is introduced in section~\ref{Methodology},
together with the theoretical tools used to obtain the $W$-pair production cross section.
Since the data-taking scheme, including the number of data-taking points, the center-of-mass (CM) energy ( $\sqrt{s}$)
of each data point,
and the allocation of the integrated luminosity, directly impacts the statistical and systematic uncertainties
of the measured $m_{W}$ and $\Gamma_{W}$, these uncertainties are studied firstly as described in section~\ref{Uncertainties}.
The investigation of the data-taking scheme and the corresponding expected precision on $m_{W}$ are presented
in section~\ref{Data taking}.

%% ============================================== Uncertainty study  =============================================
\section{Methodology and Theoretical setup}
\label{Methodology}
The cross section of the $W$-pair production around its threshold depends sensitively on the mass
and width of the $W$ boson, and the dependency can be precisely calculated in the EW theory.
Therefore by measuring the cross sections at one or more energy points around the $W$-pair threshold, the
$W$ boson mass and width can be determined by comparing the measured cross sections with the theoretical
predictions.

\begin{figure}[htbp]
	\centering
	\subfigure{
		\includegraphics[width=0.48\textwidth]{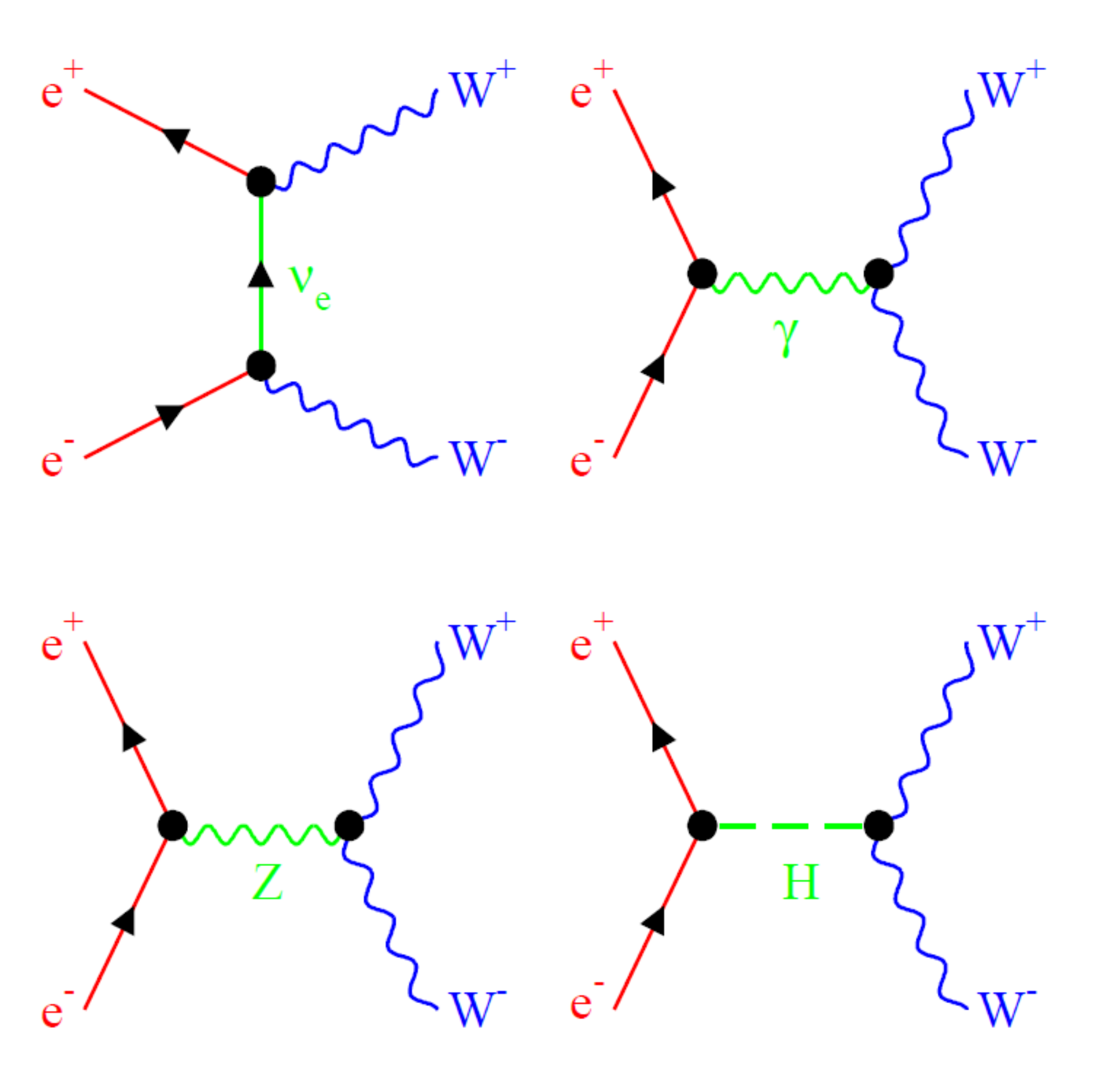}
    }
	\caption{The leading-order Feynman diagrams of $W^{+}W^{-}$
            production in $e^{+}e^{-}$ collisions. The last one is neglected
            in this study since it is
            highly suppressed due to the small electron (positron) mass.
            }
	\label{fig:Feynman_diagrams}
\end{figure}

Figure.~\ref{fig:Feynman_diagrams} shows the leading order Feynman diagrams for
$W^{+}W^{-}$ production at electron positron colliders. Due to the small electron mass,
the production of $W^{+}W^{-}$ through the Higgs boson is highly suppressed and is therefore neglected
in the discussion below. Then the Born-level matrix element of the on-shell $W^{+}W^{-}$ production can
be written as~\cite{WW_XS_1, WW_XS_2}:
\begin{equation}
\begin{aligned}
    &M = \sqrt{2}e^{2}[ M^{\nu} + M^{\gamma} + M^{Z} ]\Delta\sigma(-1)d^{J_{0}}_{\Delta\sigma,\Delta\lambda}\\
    &M^{\nu} = \frac{1}{2\sin^{2}\theta_{W}\beta}\delta_{|\Delta\sigma|,1}
    [B_{\lambda\lambda}^{\nu}-\frac{1}{1+\beta^{2}-2\beta\cos\Theta}C_{\lambda\bar{\lambda}}^{\nu}]\\
    &M^{\gamma} = -\beta\delta_{|\delta\sigma|,1}A^{\gamma}_{\lambda,\lambda}\\
    &M^{Z} = \beta[\delta_{|\Delta\sigma|,1} - \frac{1}{2\sin^{2}\theta_{W}}
    \delta_{|\Delta\sigma|,1}]\frac{s}{s-M_{Z}^{2}}A^{Z}_{\lambda\bar{\lambda}}\\
\end{aligned}
\end{equation}
where $M$ is the total amplitude of $W^{+}W^{-}$ production, $M^{\nu}$, $M^{\gamma}$,
and $M^{Z}$ are the amplitudes for the coupling channels with $\nu_e$, $\gamma$, and
$Z$, propagators, respectively;
$\Delta\sigma = \sigma - \bar{\sigma}$; $\Delta\lambda = \lambda-\bar{\lambda}$;
$\sigma$ ($\bar{\sigma}$) and
$\lambda$ ($\bar{\lambda}$) are the $z$ components of the electron (positron) and
$W^{+}$ ($W^{-}$) spins ($i.e.$ their polarization state), respectively;
$J_{0}\equiv \max(|\Delta\sigma|,\Delta\lambda)$, is the minimum angular momentum
of the system; $\beta\equiv\sqrt{1-(\frac{2m_{W}}{\sqrt{s}})^{2}}$ is the velocity
of the $W$ boson; and $\theta_{W}$ is the Weinberg weak mixing angle.

\begin{figure}[htbp]
	\centering
	\subfigure{
		\includegraphics[width=0.45\textwidth]{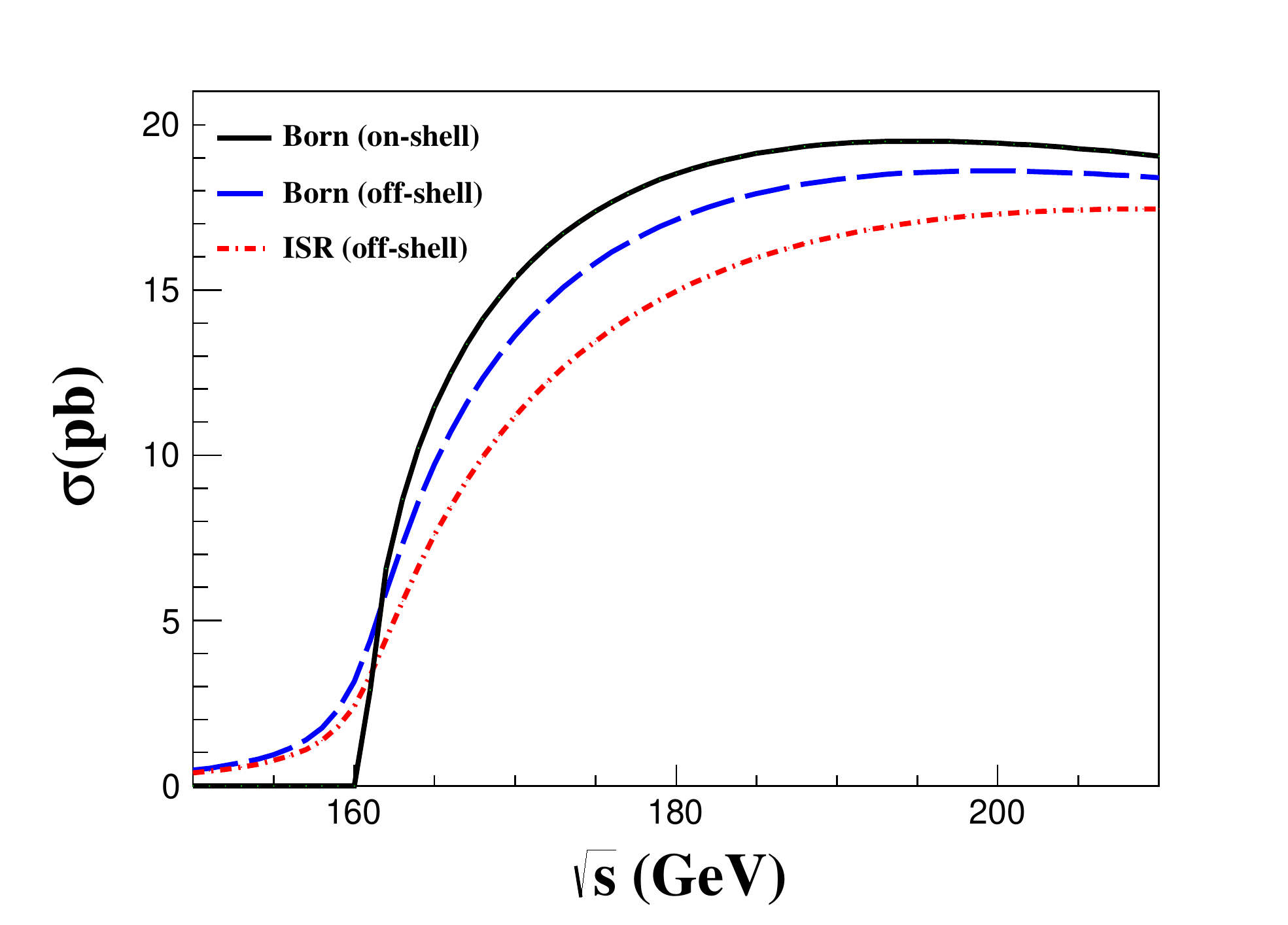}
    }
	\caption{The theoretical cross sections of $W$-pair production as functions of the CM energy of the $e^{+}e^{-}$ collisions.
    The black solid line is the Born-level cross section with the zero-width assumption, the blue dash line is the Born-level cross section
    including the finite width of the $W$ boson. The red dash-dot line is the cross section taking into account both the finite width of the
    $W$ boson and the ISR corrections. The world average values~\cite{WMass_PDG2018} of $m_W$ and $\Gamma_W$ are used in these calculations.}
	\label{fig:Gentle_XS}
\end{figure}

The production cross section of $W$-pair at $e^{+}e^{-}$ colliders,
$\sigma_{\rm{WW}}$, is calculated using the GENTLE package~\cite{Gentle} with the CC03 mode~\cite{CC11}.
%as a function of the energy ($E_{CM})$, W mass ($m_{W}$) and width ($\Gamma_{W}$).
Figure~\ref{fig:Gentle_XS} shows the cross section as functions of $\sqrt{s}$ with $m_{W}$ and $\Gamma_{W}$ fixed to their world average values: $m_{W}=80.379$~GeV and $\Gamma_{W}=2.085$~GeV~\cite{WMass_PDG2018}.
The Born-level cross sections are shown in black for a zero-width $W$ boson and in blue for a finite-width $W$ boson. The red curve includes the effects of both the finite width and the Initial State Radiation (ISR) contribution.

The goal of this study is to optimize the data-taking scheme for a fixed total integrated luminosity and
given beam parameters with their corresponding systematic uncertainties. Table~\ref{tab:configurations} summarizes
the inputs and configurations used in this study.
For comparisons, the configurations used by the FCC-ee study are also listed.

\begin{figure}[htbp]
	\subfigure{
		 \includegraphics[width=0.4\textwidth]{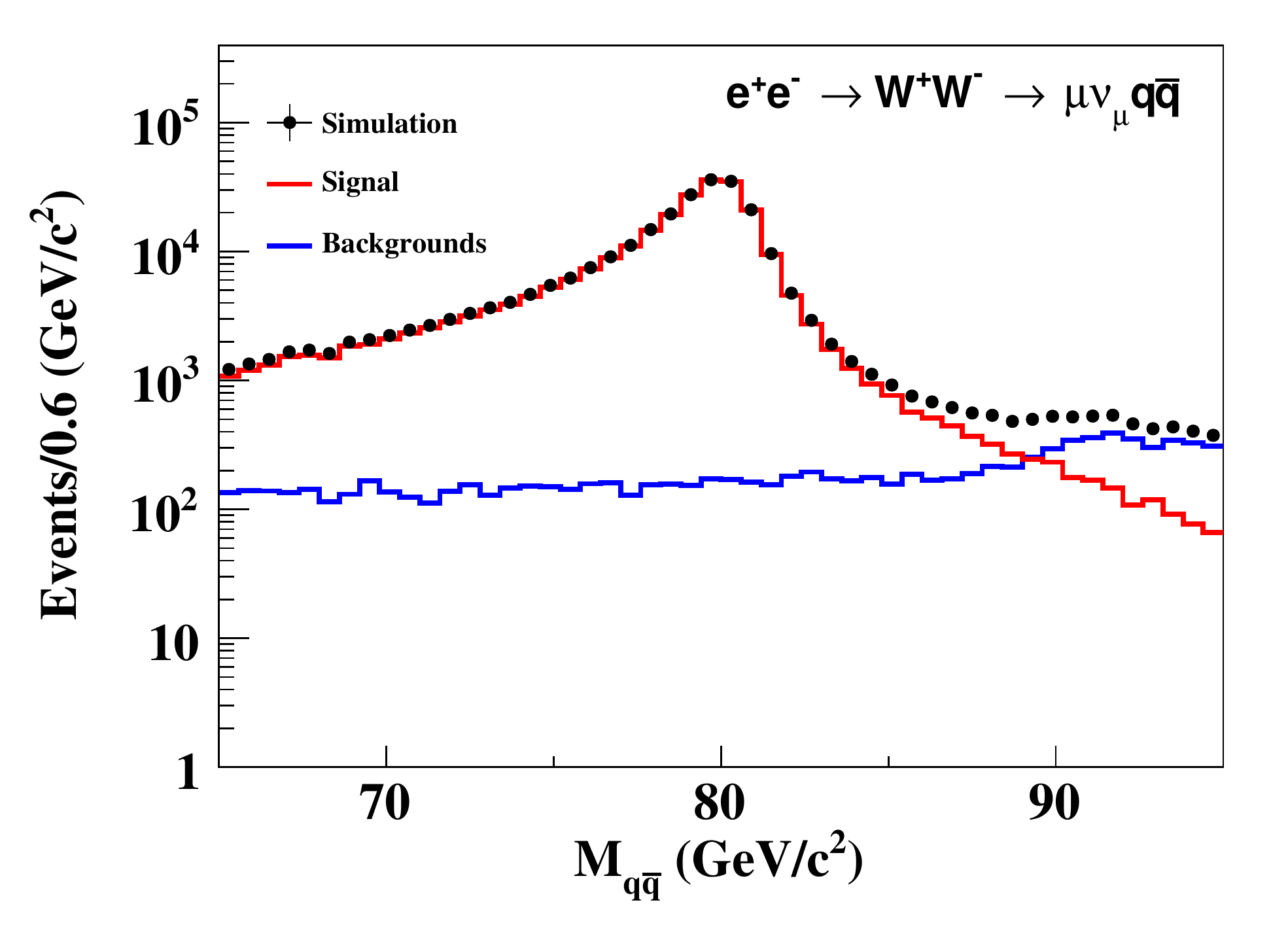}%\put(-50,110){\bf (a)}
	}
	\subfigure{
		 \includegraphics[width=0.4\textwidth]{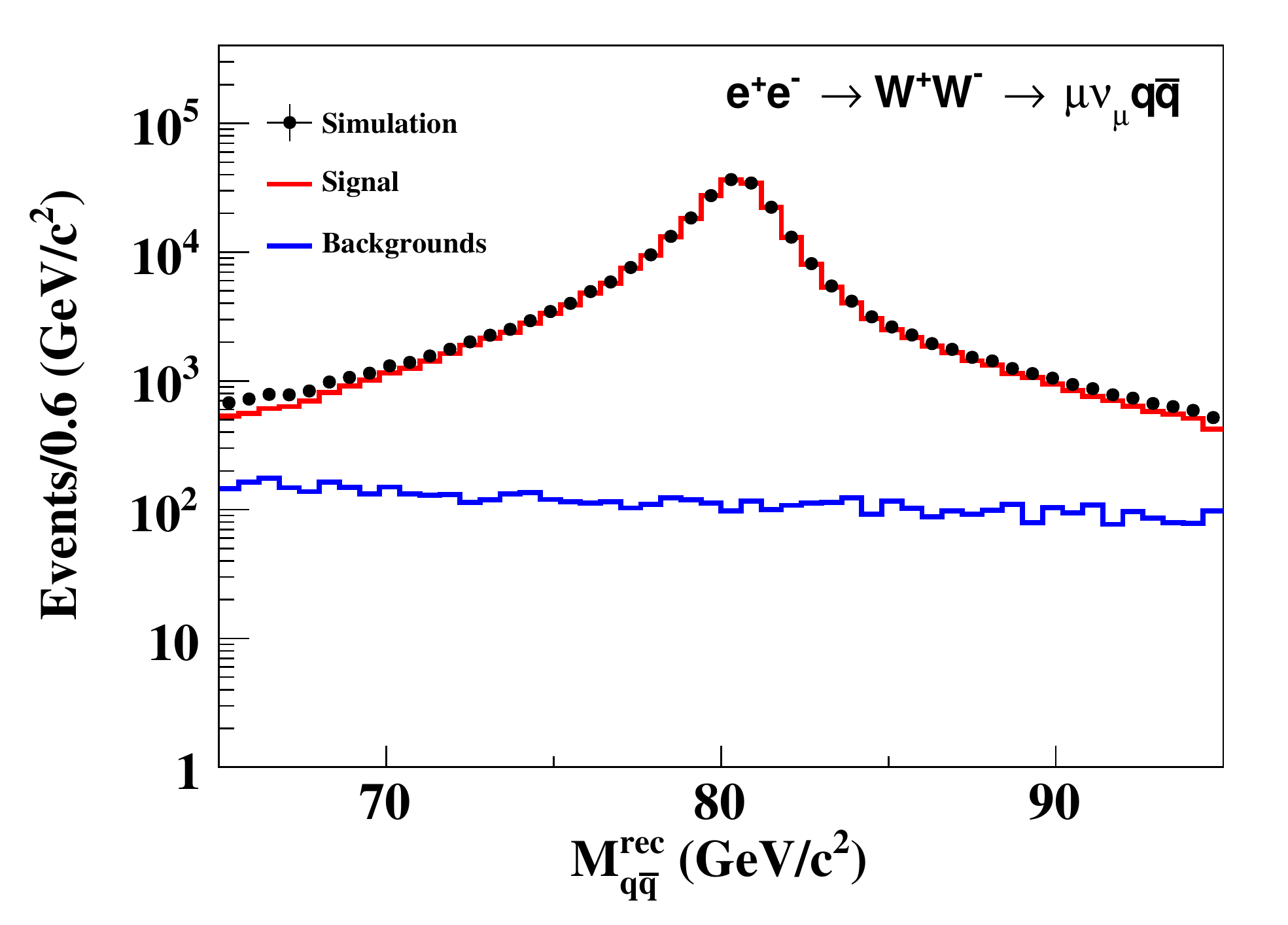}%\put(-50,110){\bf (b)}
	}
	\caption{The invariant and recoil mass of two jets of
    $e^{+}e^{-} \to W^{+}W^{-} \to \mu\nu_{\mu}q\bar{q}$ process. The black dots are the
    distributions of MC simulated SM processes, while the red and blue histograms are
    for signal and background processes, respectively.
    }
	\label{fig:eff}
\end{figure}

Among the configurations listed in Table~\ref{tab:configurations}, the mass and width of the $W$ boson are
from the PDG~\cite{WMass_PDG2018}; the total luminosity is assumed to be 3.2$~\mbox{ab}^{-1}$
expected at the CEPC in one year data-taking;
the parameters for beam energy and its spread are from the CEPC's Conceptual
Design Report~\cite{CEPC_1}; other assumptions on the systematic uncertainties are largely the same as
the ones in the FCC-ee's paper~\cite{Fcc_ee_WMass}, except for the signal selection efficiency.
To estimate the selection efficiency and purity for the $W$-pair events, the semi-leptonic
$e^{+}e^{-} \to W^{+}W^{-} \to \mu\nu_{\mu}q\bar{q}$ process is simulated  at the generator-level using the
Monte Carlo (MC) package {\sc{whizard}}~\cite{Whizard_1,Whizard_2} at $\sqrt{s}=161$~GeV.
%with all the SM backgrounds scaled to same luminosity.
The signal candidates are selected by requiring two jets, one muon.
The energy of the muon must be larger than 30~GeV. The corresponding signal selection efficiency is about 90\% with a signal purity of about 98\%.
Figure~\ref{fig:eff} shows the distributions of the invariant and recoil mass of the two selected jets.
For the pure leptonic and hadronic processes,
$e^{+}e^{-} \to W^{+}W^{-} \to l_{1}\nu_{l_{1}}l_{2}\nu_{l_{2}}/q\bar{q}Q\bar{Q}$,
signal event selections are more complex, thus the selection efficiency and
the purity are expected to be lower than those of the semi-leptonic decays.
For this study, weighted selection efficiency and purity of 80\% and 90\%, respectively, are assumed
for selecting $W$-pair events.

For the energy calibration, resonant depolarization is the most precise
method, which is successfully applied at LEP~\cite{LEP_Redipo_1,LEP_Redipo_2,LEP_Redipo_3}.
This method are proposed also by both FCC-ee~\cite{Fcc_ee,EPOL17} and CEPC~\cite{CEPC_1} for $Z$ pole and
$W$-pair threshold energy regions. And FCC-ee's study shows that the precision of 500keV
of the beam energy calibration at $W$-pair threshold can be achieved~\cite{FCC-ee_CDR}, but
require a dedicated operation mode, specific hardware elements, and carefully controlling and
monitoring of the operating conditions.

  For what concerns the energy spread, it has been shown
  in FCC-ee studies~\cite{EPOL17,EPOL17Janot} that it can be measured and monitored
  to a precision of 5\%  making use of the acollinearity distribution of
  $\sim 10^3$ dimuon events ~\cite{EPOL17}.
  For CEPC, the further study of the energy spread is in progress, so the 10\% is taken
  as the uncertainty the energy spread conservatively.

\begin{table}[htbp]
    \caption{The configurations of the data-taking assumed in this paper. Shown in the table are
    the world average values of the mass and width of the $W$ boson~\cite{WMass_PDG2018}; the total integrated luminosity and its
    relative uncertainty, $\mathcal{L}$ and $\Delta\mathcal{L}$; the means and
    uncertainties of beam energy and its spread, $E$, $E_{BS}$, $\Delta E$, and $\Delta E_{BS}$;
    the relative uncertainties of the background, and detection efficiency,
    $\Delta\sigma_{\rm{B}}$, and $\Delta\epsilon$.
    The second column is used in this study and the third one in FCC-ee's paper~\cite{Fcc_ee_WMass}.
    }
    \begin{center}
    \begin{tabular}{l|cc}
    \hline
    Configurations  & This study & FCC-ee work\\
    \hline
    $m_{W}$ (GeV) & \multicolumn{2}{c}{80.379 $\pm$ 0.012}\\
    $\Gamma_{W}$ (GeV) & \multicolumn{2}{c}{2.085 $\pm$ 0.042}\\
    \hline
    $\mathcal{L}$ ($\mbox{ab}^{-1}$) & 3.2 & 15\\
    $\sigma_{E}$ (\%) & 0.1 & 0.09 \\
    $\epsilon$ & 0.8 & 0.75 \\
    $\sigma_{\rm{B}}$ (pb) & 0.3 & 0.3 \\
    \hline
    $\Delta \sigma_{E}$ (\%) & 10 & 5\\
    $\Delta E$ (MeV) & 0.5 & 0.24\\
    $\Delta\sigma_{\rm{B}}/\sigma_{\rm{B}}$  & $10^{-3}$ & $10^{-3}$\\
    \hline
    $\Delta\mathcal{L}/\mathcal{L}$ & $10^{-4}$ & $10^{-4}$ \\
    $\Delta\epsilon/\epsilon$ & $10^{-4}$ & $10^{-4}$\\

    \hline
    \end{tabular}
    \label{tab:configurations}
    \end{center}
\end{table}

%% ============================================== Stat. uncer. =============================================
\section{Consideration on the uncertainties}
\label{Uncertainties}
Once the configurations of the data samples described above
are assumed, the data-taking scheme can be optimized.
The guideline of the optimization is to obtain the highest precision of the
mass (width) of the $W$ boson based on the fixed total integrated luminosity.
Thus the statistical and systematic uncertainties of  the $m_{W}$  and $\Gamma_{W}$ measurements
are investigated firstly, following with the estimations of the total uncertainties
of the $m_{W}$ and $\Gamma_{W}$ measurements for specific
data-taking schemes.
%Then the data taking plan is optimized to obtain the minimal $\Delta \Gamma_{W}$.

\subsection{Statistical uncertainty}
The $W$-pair cross section can be experimentally determined by counting the number of
$e^+e^- \to W^+W^-$ ($W$-pair) events. It should be noted that $W$ boson's three major
decay channels ($l\nu l\nu$, $l\nu q\bar{q}$, $q\bar{q} q\bar{q}$)
are all used to increase the statistical power. Although each channel has its own efficiencies
and background, the global analysis and the simultaneous fit could be applied
to get the total $W$-pair events.So the number of the total W pairs can be determined by
combining all various channels with all branching ratios and efficiencies taken into account.

The $W$-pair cross section at a specific CM energy point is determined by:
\begin{equation}
\label{sigma-meas}
	\sigma_{\rm{meas}} = \frac{N_{\rm{meas}}}{\mathcal{L}\epsilon} = \frac{N_{\rm{obs}}-N_{\rm{B}}}{\mathcal{L}\epsilon},
\end{equation}
where $N_{\rm{meas}}$ is the signal yield, $N_{\rm{obs}}$ and $N_{\rm{B}}$ the numbers of observed events and estimated background events, respectively,
$\mathcal{L}$ the integrated luminosity, and $\epsilon$ the signal selection efficiency.
With Eq.~\ref{sigma-meas}, the statistical uncertainty of the $\sigma_{\rm{meas}}$ can be expressed as (assuming Poisson distribution):
\begin{equation}
	\Delta\sigma_{\rm{meas}}(\rm{stat.}) \sim \frac{\sqrt{N_{\rm{meas}}}}{\mathcal{L}\epsilon} = \frac{\sqrt{\sigma_{\rm{meas}}}}{\sqrt{\mathcal{L}\epsilon P}},
	~P=\frac{\epsilon\sigma_{WW}}{\epsilon\sigma_{WW} + \epsilon_{B}\sigma_{B}},
\end{equation}
where $P$ is the signal purity of the selected sample, $\epsilon_{B}$ is the surviving rate of background events ({\em i.e.} background efficiency)
and $\sigma_{\rm{B}}$ is the total background cross section.

If the data is taken at one single energy point, the statistical sensitivities of the $W$ boson mass and width
can be obtained individually:
\begin{equation}
	\label{Stat_E}
	\begin{aligned}
		 \Delta m_{W}(\rm{stat.}) &= (\frac{\partial\sigma_{\rm{meas}}}{\partial m_{W}})^{-1}\Delta \sigma_{\rm{meas}}
					   = (\frac{\partial\sigma_{\rm{meas}}}{\partial m_{W}})^{-1}\frac{\sqrt{\sigma_{\rm{meas}}}}{\sqrt{\mathcal{L}\epsilon P}},\\
		 \Delta \Gamma_{W}(\rm{stat.}) &= (\frac{\partial\sigma_{\rm{meas}}}{\partial \Gamma_{W}})^{-1}\Delta \sigma_{\rm{meas}}
					   = (\frac{\partial\sigma_{\rm{meas}}}{\partial\Gamma_{W}})^{-1}\frac{\sqrt{\sigma_{\rm{meas}}}}{\sqrt{\mathcal{L}\epsilon P}}.
	\end{aligned}
\end{equation}

\begin{figure}[htbp]
	\subfigure{
		 \includegraphics[width=0.4\textwidth]{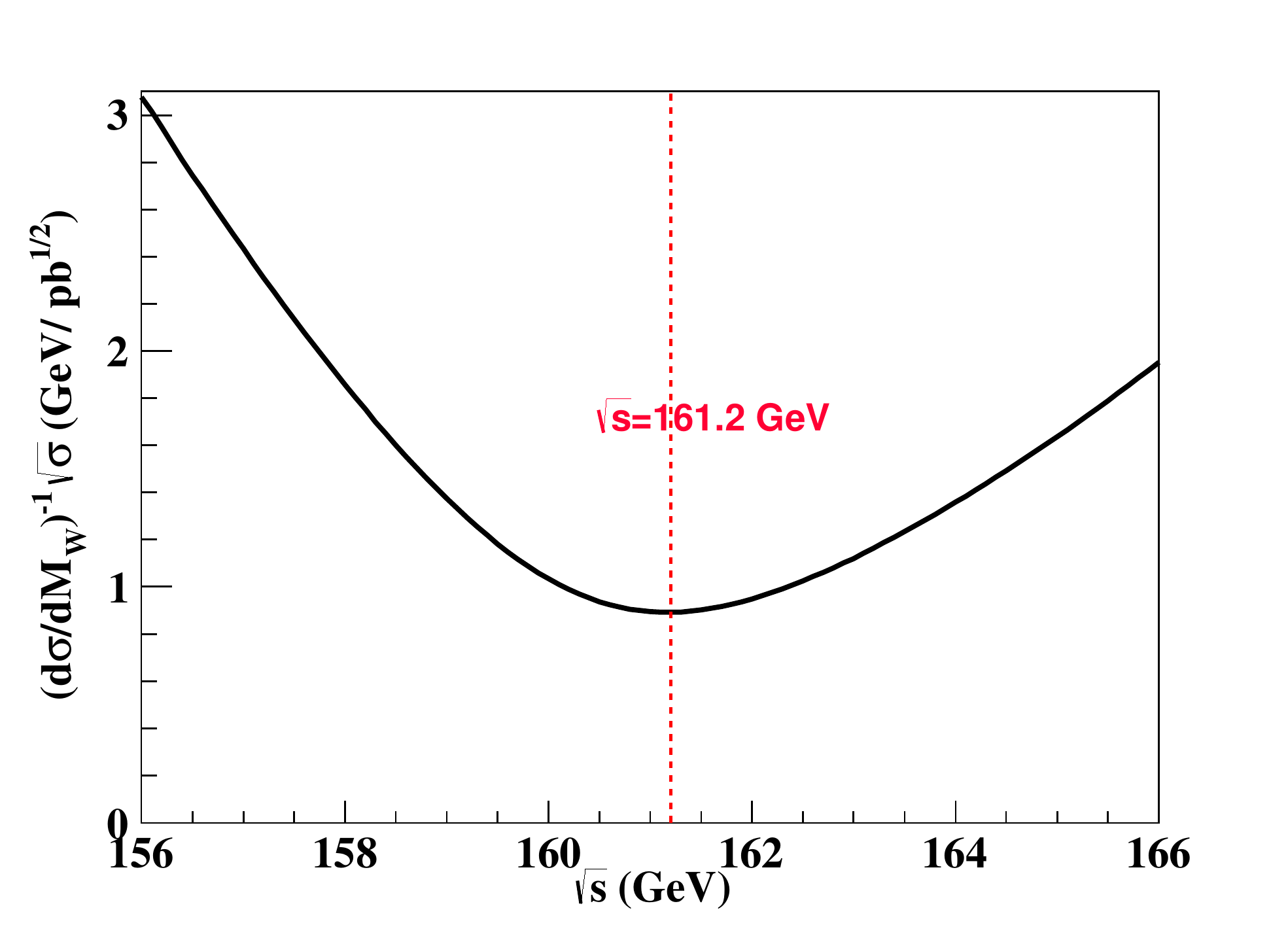}\put(-150,115){\bf (a)}
	}
	\subfigure{
		 \includegraphics[width=0.4\textwidth]{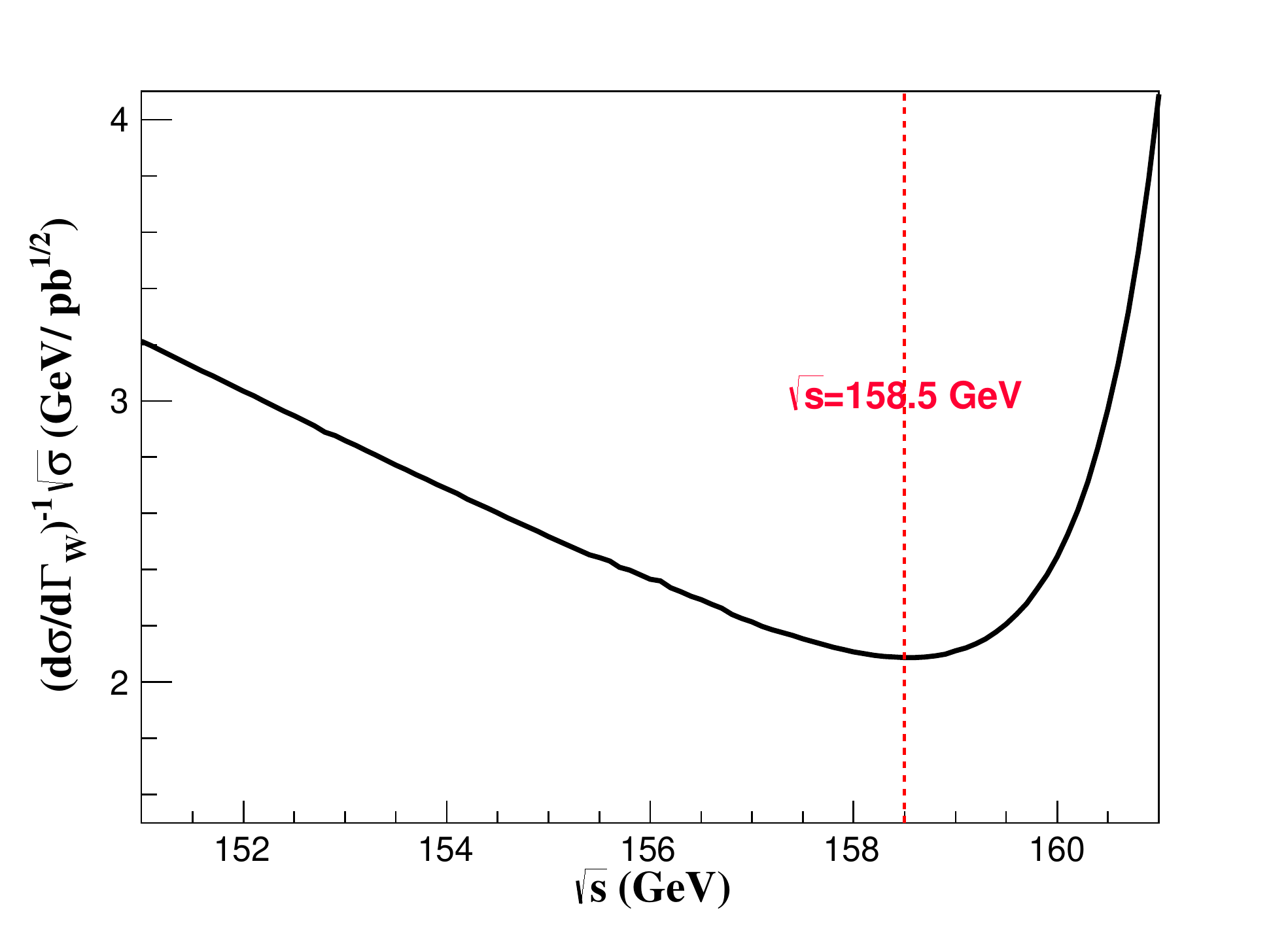}\put(-150,115){\bf (b)}
	}
	\caption{Distributions of the statistical uncertainties of $m_{W}$~(a) and
            $\Gamma_{W}$~(b) for taking data at a single energy point.}
	\label{fig:Stat_MW}
\end{figure}

Figure~\ref{fig:Stat_MW} shows the statistical uncertainties of $m_{W}$ and $\Gamma_{W}$ as  functions of $\sqrt{s}$ of the data-taking. The distributions show minimal statistical uncertainties for $m_{W}$ and $\Gamma_{W}$, but at two different $\sqrt{s}$ values. Please note, however, only one of them can be determined
at one single data point, with the another one fixed to the world averaged value~\cite{WMass_PDG2018}.

For taking data at more than one energy point, $m_{W}$ and $\Gamma_{W}$ can be measured simultaneously.
The statistical uncertainties can be obtained by  the covariance matrix,
which is the inverse of the second-order derivative matrix of the log-likelihood or $\chi^{2}$ function with
respect to its free parameters, usually evaluated at their best values (the function minimum).
The minimum $\chi^{2}$ method is used in this study and the $\chi^{2}$ is constructed
as:
\begin{equation}
	\label{Chi2_1}
	\chi^{2} = \sum_{i}\frac{(N_{\rm{fit}}^{i} - N_{WW}^{i})^{2}}{N_{WW}^{i}}~,
\end{equation}
which is minimized using the {\sc{minuit}} package~\cite{Minuit}.

Therefore the covariance matrix can be written as:
%{\color{red} the matrix is necessary? Only used here? check a partial operation in 2nd matrix}
\begin{equation}
\begin{large}
\begin{aligned}
	\label{EMatrix}
	V &= \frac{1}{2}
		\begin{bmatrix}[cc]
			\frac{\partial^{2}\chi2}{\partial m_{W}^{2}} & \frac{\partial^{2}\chi2}{\partial m_{W}\partial\Gamma_{W}}\\
			\frac{\partial^{2}\chi2}{\partial m_{W}\partial\Gamma_{W}} & \frac{\partial^{2}\chi2}{\partial m_{\Gamma_{W}}^{2}}\\
		\end{bmatrix}^{-1}\\
	  &= \begin{bmatrix}[cc]
      \sum\limits_{i}\frac{1}{(\Delta\sigma_{\rm{meas}}^{i})^{2}}
	  \begin{bmatrix}[cc]
		  (\frac{\partial \sigma^{i}}{\partial m_{W}})^{2} &
		  \frac{\partial \sigma^{i}}{\partial m_{W}}\frac{\partial \sigma^{i}}{\partial \Gamma_{W}} \\
		  \frac{\partial \sigma^{i}}{\partial m_{W}}\frac{\partial \sigma^{i}}{\partial \Gamma_{W}} &
		  (\frac{\partial \sigma^{i}}{\partial \Gamma_{W}})^{2} \\
	  \end{bmatrix}
      \end{bmatrix}^{-1}.
\end{aligned}
\end{large}
\end{equation}

The diagonal elements of the second-order derivative matrix, are de-coupled from other parameter(s),
but when the matrix is inverted, the diagonal elements of the inverse contain contributions from
all the elements of the second derivative matrix.
When the number of fit parameters is reduced to one,  Eq.~\ref{EMatrix} is simplified to Eq.~\ref{Stat_E}.

\begin{figure}[htbp]
	\subfigure{
		 \includegraphics[width=0.4\textwidth]{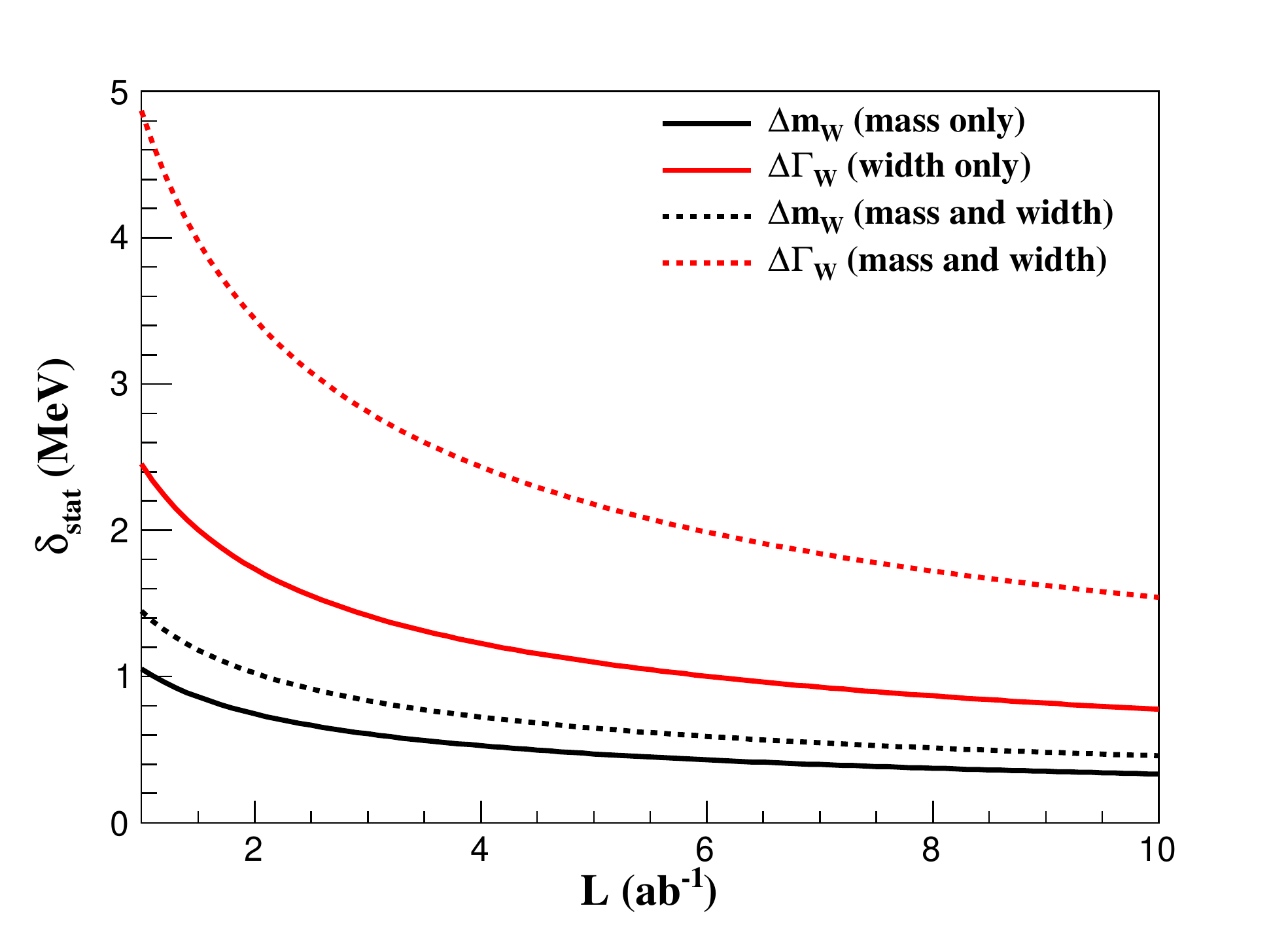}\put(-160,110){\bf (a)}
	}
	\subfigure{
		 \includegraphics[width=0.4\textwidth]{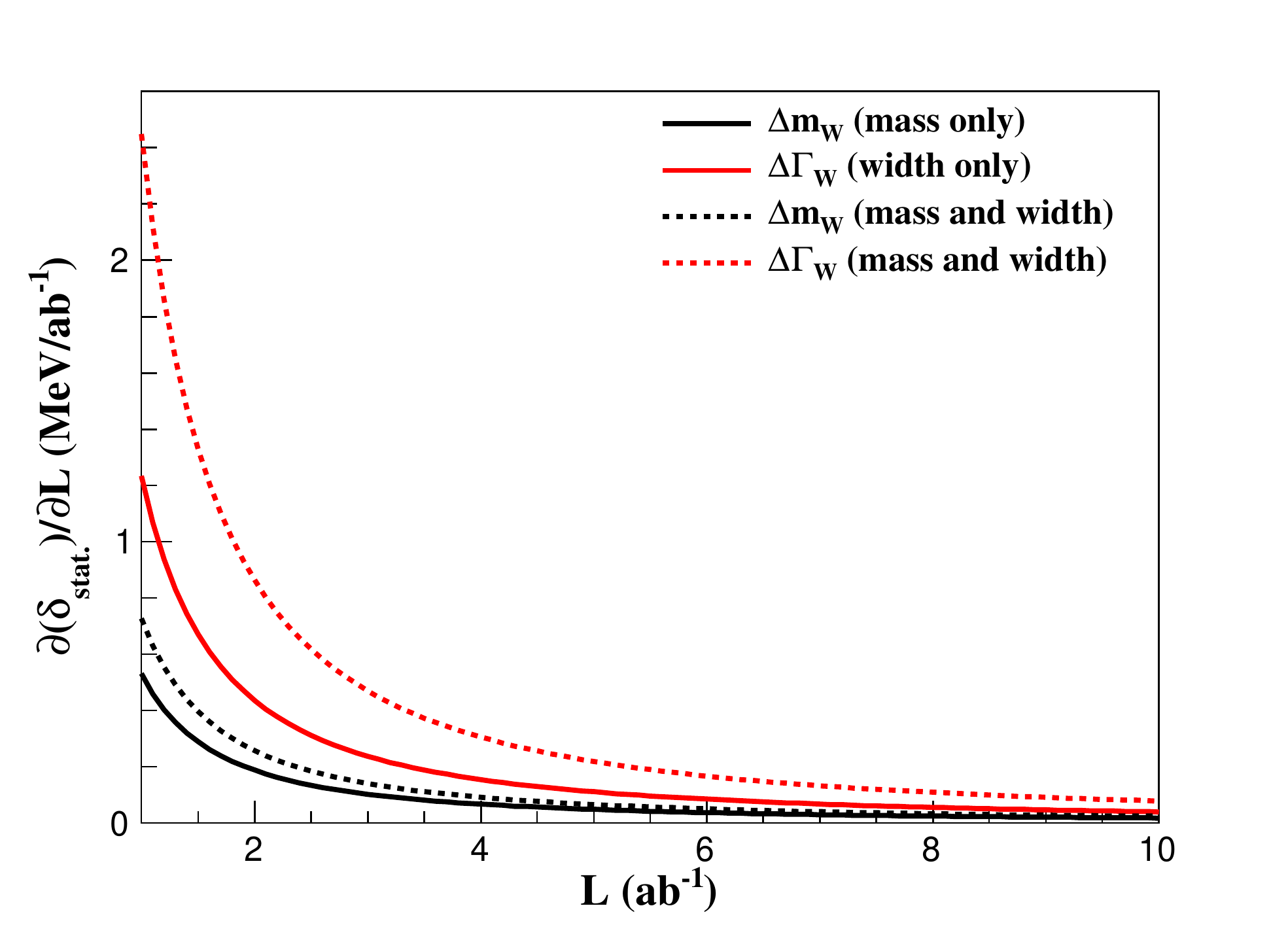}\put(-160,110){\bf (b)}
	}
	\caption{(a)The dependence of the statistical uncertainties of the measured results
        on the statistic of data.
        (b)The decline rate of the statistical uncertainty to luminosity.
        The black (red) solid line shows the result for measuring the $m_{W}$ ($\Gamma_{W}$) only,
        and the black and red dots show the results for measuring the $m_{W}$ and $\Gamma_{W}$ simultaneously.
        The energy 162.5~GeV is used for $m_{W}$ and 158.5~GeV is used for $\Gamma_{W}$, and
        they are both used when the mass and width are fitted simultaneously.}
	\label{Precision_L}
\end{figure}

Fig.~\ref{Precision_L}(a) shows that
the dependence of the precision of $m_{W}$ and/or $\Gamma_{W}$  are in inversely proportional
to integrated luminosity, which is consistent with the Eq.~\ref{Stat_E} and~\ref{EMatrix}.
The derivatives of the statistical uncertainties are shown in
the Fig.~\ref{Precision_L}(b), and it become almost stable when luminosity is greater than 6~ab$^{-1}$.

%% ============================================== Sys.  uncer. =============================================
\subsection{Systematic uncertainties}
\label{Section_sys}
Since the $W$ boson mass and width are determined by comparing the measured cross section(s)
of $W$-pair with the theoretical prediction(s), there are various sources could contaminate the measured precisions,
which can be separated into two categories: 1) The "X-value" (abscissa of Fig.~\ref{fig:Gentle_XS})
uncertainties, such as the beam energy calibration ($E$) and
the beam energy spread ($\sigma_{E}$) measurement. Generally speaking, there will be some dedicated approaches
to measure $E$ and $\sigma_{E}$, and the uncertainties after the measurements are defined as $\Delta E$ and $\Delta \sigma_{E}$,
respectively;
2) The "Y-value" (ordinate of Fig.~\ref{fig:Gentle_XS}) or "yield uncertainties", which are affected by
the integrated luminosity, the selection efficiency and the background determinations.

\subsubsection{"X-value" uncertainties}
The energy and energy spread of each beam are associated with the accelerator performance, and
their uncertainties are treated as point-to-point, which means that these uncertainties of different
data points are independent with each other. The uncertainties of energy and energy spread of each
beam are both assumed to follow Gaussian distribution, $E=G(E_{0}, \Delta E)$ and
$\sigma_{E}=G(\sigma_{E}^{0}, \Delta \sigma_{E})$, where $E_{0}$ and $\sigma_{E}^{0}$ are the nominal values
for the energy and its spread, respectively.

Take the energy spread into account, the measured cross section at a specific energy point, $E_{0}$, reads:
\begin{equation}
	\begin{aligned}
    \label{Eq_Espread}
		\sigma_{WW}(E_{0}, \sigma_{E}^{0}) &= \int \sigma_{WW}(E)\times G(E_{0},\sigma_{E}^{0})dE \\
				 &= \int \sigma_{WW}(E)\times \frac{1}{\sqrt{2\pi}\sigma_{E}^{0}}e^{\frac{-(E^{0}-E)^{2}}{2{\sigma_{E}^{0}}^{2}}}dE.
	\end{aligned}
\end{equation}
When both  $\Delta E$ and $\Delta \sigma_{E}$ are considered, the $\sigma_{WW}$ becomes:
\begin{equation}
	\sigma_{WW}(E_{0}, \sigma_{E}^{0}) = \int \sigma_{WW}(E^{'})\times \frac{1}{\sqrt{2\pi}\sigma_{E}}e^{\frac{-(E-E^{'})^{2}}{2{\sigma_{E}}^{2}}}dE^{'},
\end{equation}

%================================= For the uncertainty of energy calibration ====================================
The $\Delta m_{W}$ associated with the $\Delta E$ can be written as
\begin{equation}
    \label{Eq_S_E}
    \Delta m_{W}(\Delta E) = \frac{\partial m_{W}}{\partial E}\cdot \Delta E
                           = \frac{\partial m_{W}}{\partial\sigma_{WW}}\cdot
                                      \frac{\partial \sigma_{WW}}{\partial E}\cdot \Delta E .
\end{equation}

\begin{figure}[htbp]
	\subfigure{
		 \includegraphics[width=0.45\textwidth]{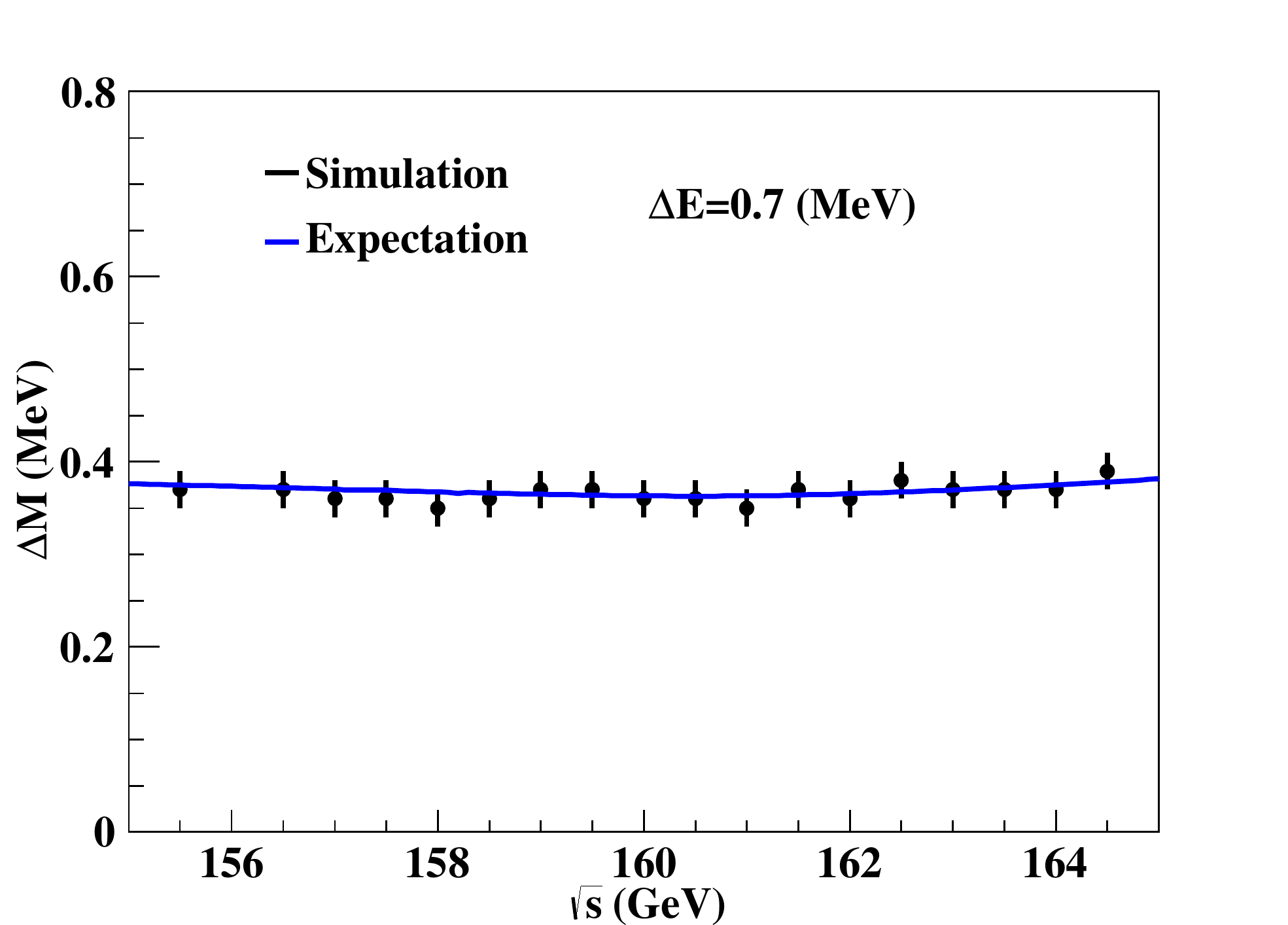}
	}
	\caption{(The dependence of $\Delta m_{W}$ on the energy with a the $\Delta E=0.7$~MeV.
            The black dots with error bars are the simulations results and the blue curve
            is the numerical result of Eq.~\ref{Eq_S_E}}
	\label{fig:SysE}
\end{figure}

Figure~\ref{fig:SysE} shows the dependence of the uncertainty of $m_{W}$ on the $\Delta E$,
with $\Delta E=0.7$~MeV (since the two beam energies are thought to be independent,
the uncertainty of the beam energy is 0.5~MeV, and 0.7~MeV for the total CM energy).
The black dots with error bars are the simulations results and the blue curve is the numerical result from Eq.~\ref{Eq_S_E},
which are consistent with each other. It  can be seen that the $\Delta m_{W}$ associated with the $\Delta E$
almost insensitive to the energy from 155~GeV to 165~GeV, which indicates that this uncertainty can be estimated separately
with the optimization of the data-taking strategy.

%================================= For the uncertainty of energy spread ====================================
\begin{figure}[htbp]
	\centering
	\subfigure{
		 \includegraphics[width=0.45\textwidth]{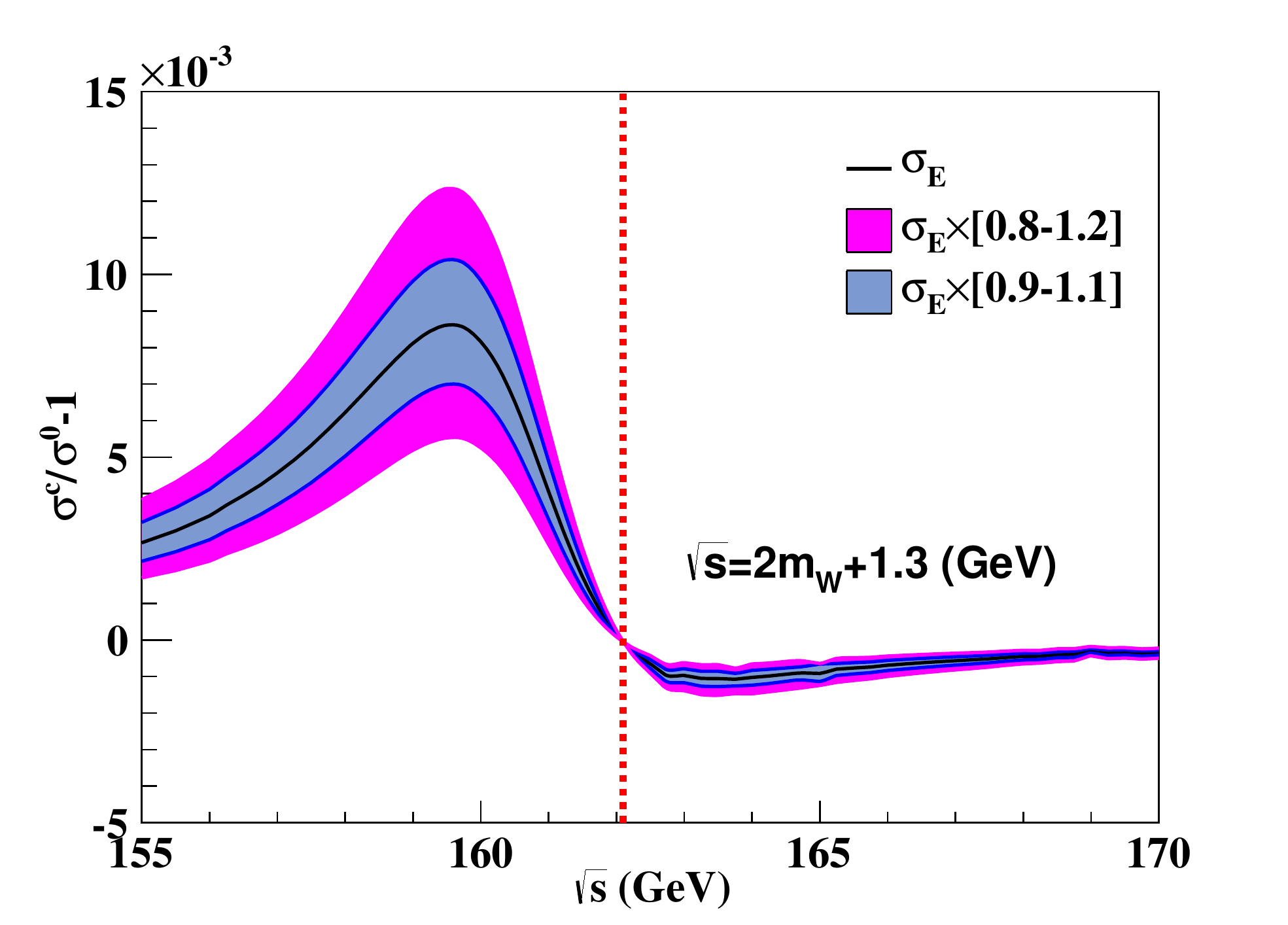}
	}
	\caption{The distribution of the ratio between the cross sections with different
        $\sigma_{E}$ and the one without the energy spread. The central curve corresponds to the
        prediction obtained	with $\sigma_{E}=0.1\%$ (relative value), which is the design
        value of the CEPC. Purple and blue bands show the ratio curves obtained varying the nominal
        $\sigma_{E}$ with [0.8, 1.2]$\sigma_{E}$.}
	\label{fig:SysBES}
\end{figure}

The distributions of $W$-pair cross section with different energy spreads are shown in
Fig.~\ref{fig:SysBES}, whose Y-axis is the ratio between the cross sections with different
$\delta_{E}$ and the nominal one without the spread.
We can see that the dependence of cross section on the beam energy spread intersects at a point
at $E\approx 2m_{W} + 1.3$~GeV, which means that the cross section is insensitive to the beam energy spread
in the vicinity of  this specific energy point. So the effects of the energy spread and its uncertainty to the cross section are
both can be neglected here.
Analytic way to consider the effect of the energy spread can be performed using the Taylor
expansion of the $\sigma_{WW}$~\cite{FCCW18_2Azzurri}, which reads
\begin{equation}
    \label{TaylorExp}
    \begin{aligned}
    \sigma_{WW}(E_{0}) &= \sigma_{WW}(E_{0}) + \frac{d\sigma_{WW}}{dE}(E-E_{0}) + \\
                          &\frac{1}{2}\frac{d^{2}\sigma_{WW}}{dE^{2}}(E-E_{0})^{2}+ ... + \frac{1}{n!}\frac{d^{n}\sigma_{WW}}{dE^{n}}(E-E_{0})^{n}.
    \end{aligned}
\end{equation}

With the above expansion, the Eq.~\ref{Eq_Espread} becomes
\begin{equation}
    \label{Eq_Taylor_result}
    \begin{aligned}
		\sigma_{WW}(E_{0}, \sigma_{E}^{0}) &= \sigma_{WW}(E_{0}) + \frac{1}{2}\frac{d^{2}\sigma_{WW}}{dE^{2}}{(\sigma_{E}^{0}\cdot E_{0})}^{2}\\
            &\frac{1}{8}\frac{d^{4}\sigma_{WW}}{dE^{4}}{(\sigma_{E}^{0}\cdot E_{0})}^{4} + ... .
    \end{aligned}
\end{equation}

The variation of the cross section is
\begin{equation}
    \Delta \sigma_{WW}(E_{0}, \sigma_{E}^{0}) = \frac{1}{2}\frac{d^{2}\sigma_{WW}}{dE^{2}}{(\sigma_{E}^{0}\cdot E_{0})}^{2}.
\end{equation}

The third item of Eq.~\ref{Eq_Taylor_result} is about two orders of magnitude smaller than the second one, therefore
the high order items can be neglected safely. So the effect of the uncertainty of energy spread on the $m_{W}$ can be expressed as

\begin{equation}
    \label{Eq_Spread_err}
    \Delta m_{W} (\Delta \sigma_{E}^{0}) = \Delta \sigma_{E}^{0}\cdot\frac{\partial m_{W}}{\partial \sigma_{WW}}
                            \cdot\frac{d^{2}\sigma_{WW}}{dE^{2}}\cdot {(\sigma_{E}^{0}\cdot E_{0})}^{2}
\end{equation}

With $\sigma_{E}^{0}=0.001$, $\Delta \sigma_{E}^{0}=0.1$, $\frac{\partial m_{W}}{\partial \sigma_{WW}}=0.48$
and $\frac{d^{2}\sigma_{WW}}{dE^{2}}=0.16$ at 161.2~GeV, $\Delta m_{W}$ associated with $\Delta \sigma_{E}^{0}$
is about 0.2~MeV, which is consistent with the result obtained by simulation.

\subsubsection{"Y-value" uncertainties}
\label{Section_CorrSys}
From the Eq.~\ref{sigma-meas} we can see that the signal yield (number of $W^+W^-$ events), is affected by the
uncertainties related to the luminosity, efficiency, and background in different ways, which can be shown
from the error propagation:
\begin{equation}
    \Delta \sigma_{meas}^{2} = \sigma_{meas}^{2}(\Delta \mathcal{L}^{2} + \Delta\epsilon^{2}) +
                            \frac{1}{\mathcal{L}^{2}\epsilon^{2}}\Delta N_{B}^{2},
\end{equation}
where $\Delta\mathcal{L}$ and $\epsilon$ are the relative uncertainties of luminosity and efficiency,
and $\Delta N_{B}$ is the uncertainty of background.

%================================= For the uncertainty of backgrounds ====================================
For the $WW$ production above their threshold, the potential main background processes include
$e^{+}e^{-}\to Z^{0}/\gamma \to q\bar{q}$, $e^+e^- \to Z^{0}e^+e^-$, $e^+e^- \to Z^{0}Z^{0}$, $e^+e^- \to We\bar{\nu_e}$,
and $e^+e^- \to \tau^+\tau^-$~\cite{ALEPH, DELPHI, L3, OPAL}.
The effect of the backgrounds has  two parts, the statistical fluctuation and
the uncertainty of the theoretical predictions of their cross sections.
The effective background cross section is set as 0.3pb in this study, which is consistent with LEP2's
result~\cite{ALEPH, DELPHI, L3, OPAL} and the one in FCC-ee's work~\cite{Fcc_ee}.
The contribution of the statistical uncertainty of background to $m_{W}$ is
\begin{equation}
    \label{Eq_S_bkg}
    \Delta m_{W}(\Delta N_{\rm{B}}) = \frac{\partial m_{W}}{\partial\sigma_{WW}}\cdot
    \frac{\sqrt{\mathcal{L}\epsilon_{\rm{B}}\sigma_{\rm{B}}}}{\mathcal{L}\epsilon},
\end{equation}

\begin{equation}
    \label{Eq_T_bkg}
    \Delta m_{W}(\Delta\sigma_{\rm{B}}) = \frac{\partial m_{W}}{\partial\sigma_{WW}}\cdot
    \frac{\mathcal{L}\epsilon_{\rm{B}}\sigma_{\rm{B}}}{\mathcal{L}\epsilon}\cdot\Delta\sigma_{\rm{B}},
\end{equation}

where $\epsilon_{\rm{B}}$ and $\sigma_{B}$ are the selection efficiency and cross section of
backgrounds, respectively, and their product is the effective background cross section, and $\Delta\sigma_{\rm{B}}$
is the relative uncertainties of the background cross section.
The ratio of them can be written as

\begin{equation}
    \label{Eq_R_bkg}
    R = \frac{\Delta m_{W}(\Delta N_{\rm{B}})}{\Delta m_{W}(\Delta\sigma_{\rm{B}})}
      = \frac{\sqrt{\mathcal{L}\epsilon_{\rm{B}}\sigma_{\rm{B}}}}{\mathcal{L}\epsilon_{\rm{B}}\sigma_{\rm{B}}\cdot \Delta\sigma_{\rm{\rm{B}}}}
      = \frac{1}{\Delta\sigma_{\rm{B}}\sqrt{\mathcal{L}\epsilon_{\rm{B}}\sigma_{\rm{B}}}},
\end{equation}

With $\mathcal{L}=3.2$~ab$^{-1}$, $\Delta\sigma_{\rm{B}}=10^{-3}$,
and $\epsilon_{\rm{B}}\sigma_{\rm{B}}=0.3$~pb at 161.2~GeV, the corresponding $R\sim1$ and
the $\Delta m_{W} (\Delta \sigma_{B})$ is about $0.2$MeV. The contributions of $\Delta m_{W} (\Delta N_{B})$
have already been considered by embodying in the product of the efficiency and purity as shown in Eq.~\ref{Stat_E},
which is a simply dilution of the statistical power of data sample. It should be noted that the input for
$\Delta \sigma_{B}$ used in this work is at $10^{-3}$ level, which is comparable with FCC-ee's work~\cite{Fcc_ee_WMass}.
But this assumption is quite challenging for the current knowledge about the background, especially for the
hadronic processes. Based on this spot, the background could be studied using the data-based method,
such as data samples collected at the $Z$ pole or below the $W$-pair threshold to calibrate the background.
If systematic uncertainty of background could be controlled at same level as its statistical part,
this uncertainty will not be limiting the precision of the measured $m_{W}$.

The uncertainties of luminosity and efficiency affect the cross section in same way,
so we define the combined uncertainty, $\delta_{c} \equiv\sqrt{\Delta\mathcal{L}^{2}+\Delta\epsilon^{2}}=1.4\cdot 10^{-4}$,
to consider their total contributions of these two items. The $\Delta m_{W}$ associated with $\delta_{c}$ is
\begin{equation}
  \label{Eq_ErrLum}
  \Delta m_{W} (\delta_{c}) = \frac{\partial m_{W}}{\partial\sigma_{WW}}\sigma_{WW}\cdot\delta_{c}~.
\end{equation}
With Eq.~\ref{Eq_ErrLum}, one can obtain the $\Delta m_{W}$ of a specific energy point, and the similar result
can be applied for $W$ width.

When there is more than one energy point, the uncertainties of luminosity and efficiency have often been treated
as the non point-to-point ones in experiment. Since the same detector, signal model, and
theoretical calculation (Bhabha process for determining the luminosity) are used for all the energy points.
The cross sections of $W$-pair production at different energy points are expected to vary
in same unknown direction and in similar relative amount simultaneously, which means that
these uncertainties of different data points are correlated.

There are two common to consider $\delta_{c}$:
\begin{itemize}
    \item[1)]  Gaussian case: $\delta_{c}$ is assumed to follow Gaussian distribution, which means that
        the cross section can be written as
        \begin{equation}
             \sigma_{WW} = G(\sigma_{WW}^{0}, \sigma_{WW}^{0}\cdot\delta_{c}),
        \end{equation}
        where $\sigma_{WW}^{0}$ is the nominal value.
    \item[2)]
        Non-Gaussian case: the $\delta_{c}$ is treated as a fixed value, so the cross section is
        \begin{equation}
              \sigma_{WW} = \sigma_{WW}^{0}\cdot(1+\delta_{c}).
        \end{equation}
\end{itemize}

\begin{figure}[htbp]
	\subfigure{
		 \includegraphics[width=0.4\textwidth]{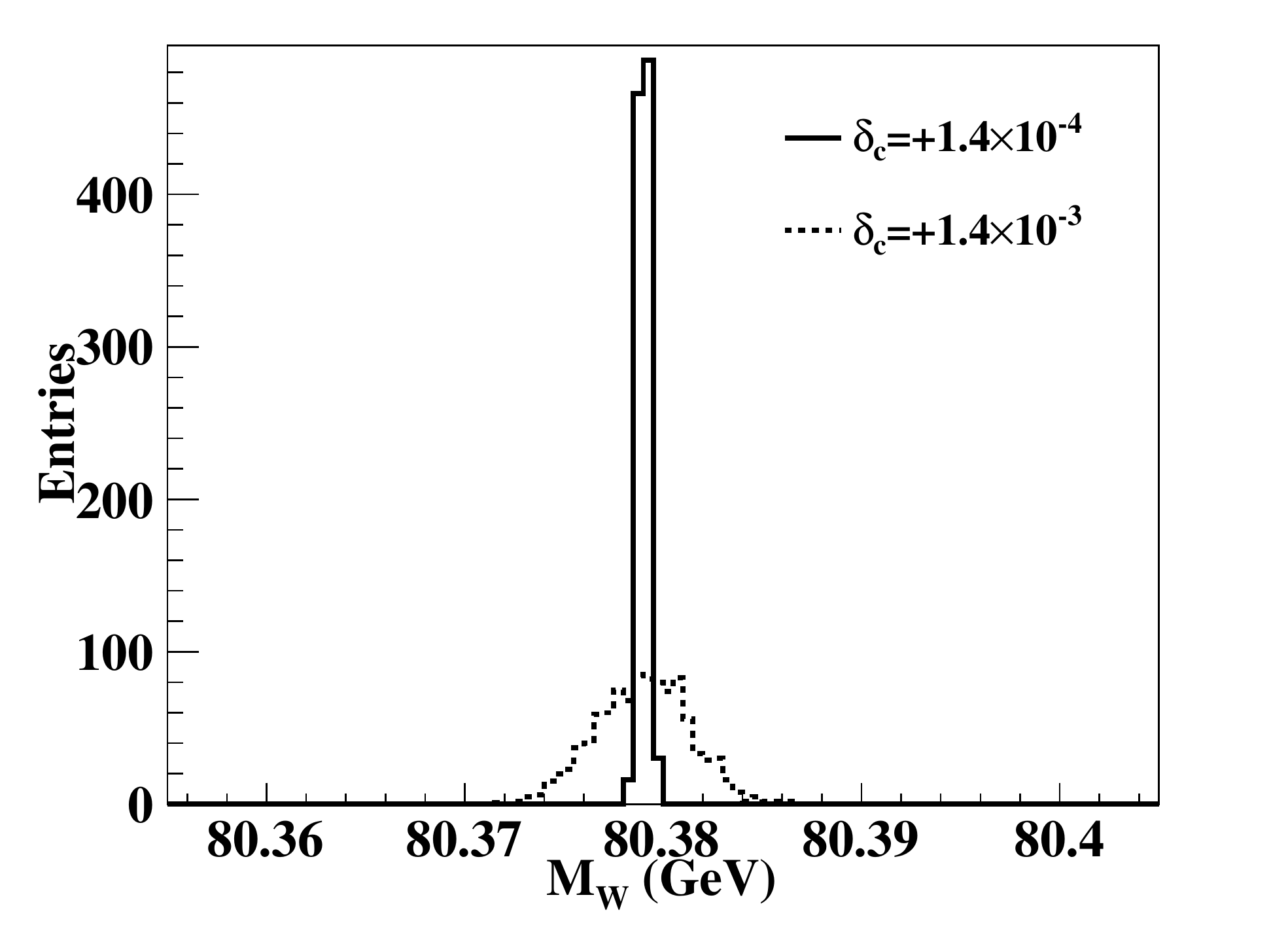}\put(-150,105){\bf (a)}
	}
	\subfigure{
		 \includegraphics[width=0.4\textwidth]{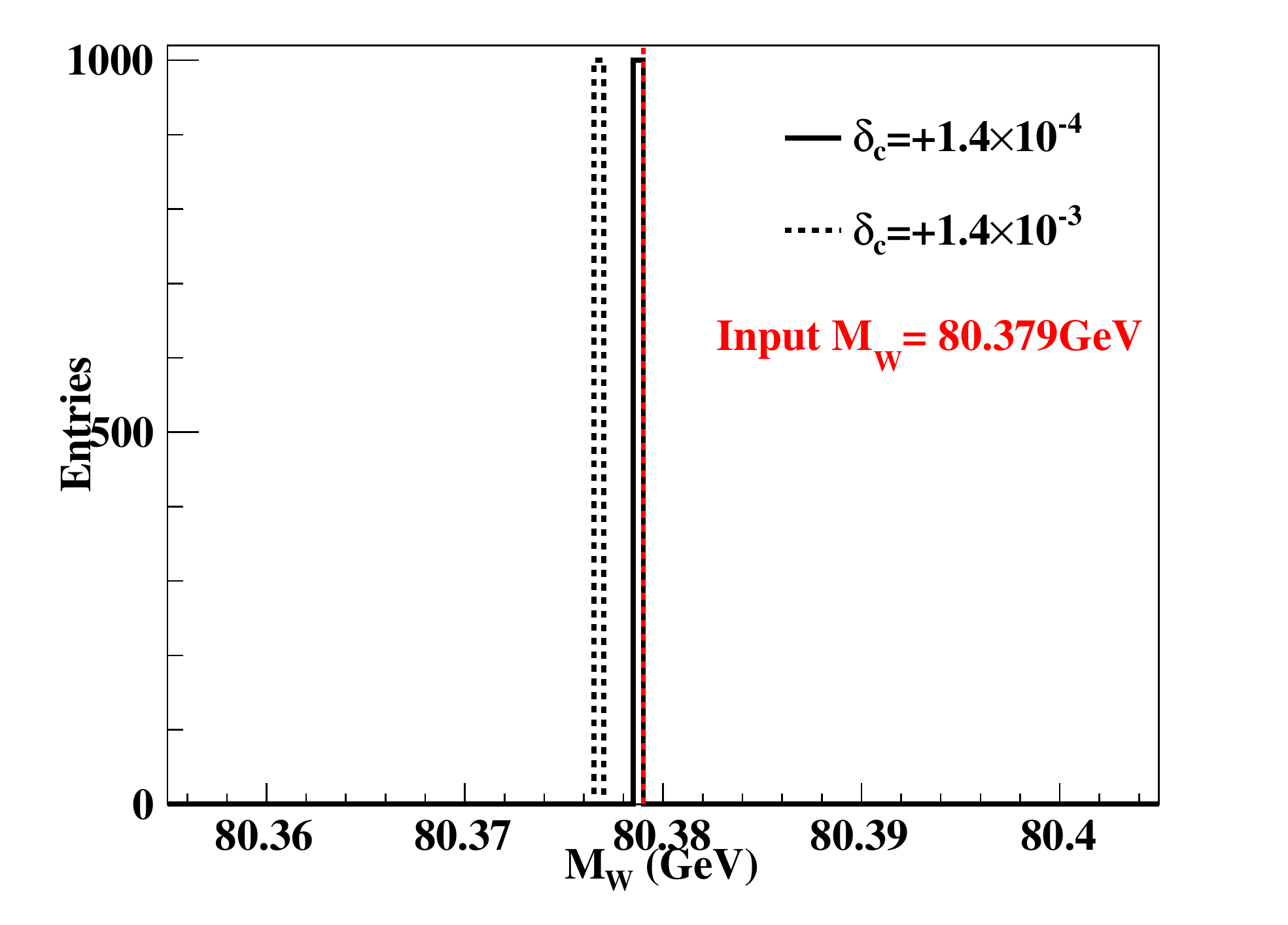}\put(-150,105){\bf (b)}
	}
	\caption{The simulation results of $m_{W}$ for the Gaussian (a) and non-Gaussian (b) cases,
        with the input $m_{W}$ of 80.379~GeV. The black solid and dash lines are for
        $\delta_{c}=+1.4\cdot10^{-4}$ and $\delta_{c}=+1.4\cdot10^{-3}$ at 161.2~GeV, respectively.}
	\label{fig:SysCorr_nomal}
\end{figure}

For the Gaussian case, the measured $m_{W}$ follows the Gaussian distribution as well, and its
standard deviation is $\Delta m_{W}$.
Figure~\ref{fig:SysCorr_nomal}(a) shows the simulation results  with $\delta_{c}= +1.4\cdot10^{-4}$
and $+1.4\cdot10^{-3}$ at 161.2~GeV (the $\delta_{c}$ is enlarged 10 times for demonstration).
The fitted $\Delta m_{W}$ are 0.24 and 2.4~MeV, respectively, which are consistent with
the direct calculations from the Eq.~\ref{Eq_ErrLum}.

For the non-Gaussian case, the situation is different.
The measured cross section will be changed of $\sigma_{WW}\cdot\delta_{c}$,
and the $\Delta m_{W}$ is turned to be the shift now, as shown in Fig.~\ref{fig:SysCorr_nomal}(b).
We can see that the fitted $m_{W}$ is shifted to left with positive inputs for $\delta_{c}$,
since the $\partial m_{W}/\partial \sigma_{WW}$ is negative at this energy.
This shifts becomes significant with the increasing $\delta_{c}$, so the correlation should be
taken into account to reduce the contribution from $\delta_{c}$, especially for the non-Gaussian case.

In general, there are several ways
to consider the correlation between multiple energy points in experiment, such as the covariance matrix
and scale factor methods~\cite{CELLO_R,ConvMatrix}. These two methods are discussed and compared in
the Refs.~\cite{Chisq_Equi,Four_Chisq,Further_Chisq,Sys_Chi2} and the latter is used in this work,
with the $\chi^{2}$ constructed as
\begin{equation}
	\label{Chi2_Corr}
	\chi^{2} = \sum_{i}\frac{(N_{\rm{meas}}^{i} - h\cdot N_{\rm{fit}^{i}} )^{2}}{\delta_{i}^{2}} + \frac{(h-1)^{2}}{\delta_{c}^{2}},
\end{equation}
where $\delta_{i}$ is the combination of the statistical and uncorrelated systematic uncertainties,
$\delta_{c}$ is the total relative correlated systematic uncertainty,
$h$ is a free parameters and $(h-1)$ represents the potential shift of the measurement.

\begin{figure}[htbp]
	\subfigure{
		 \includegraphics[width=0.4\textwidth]{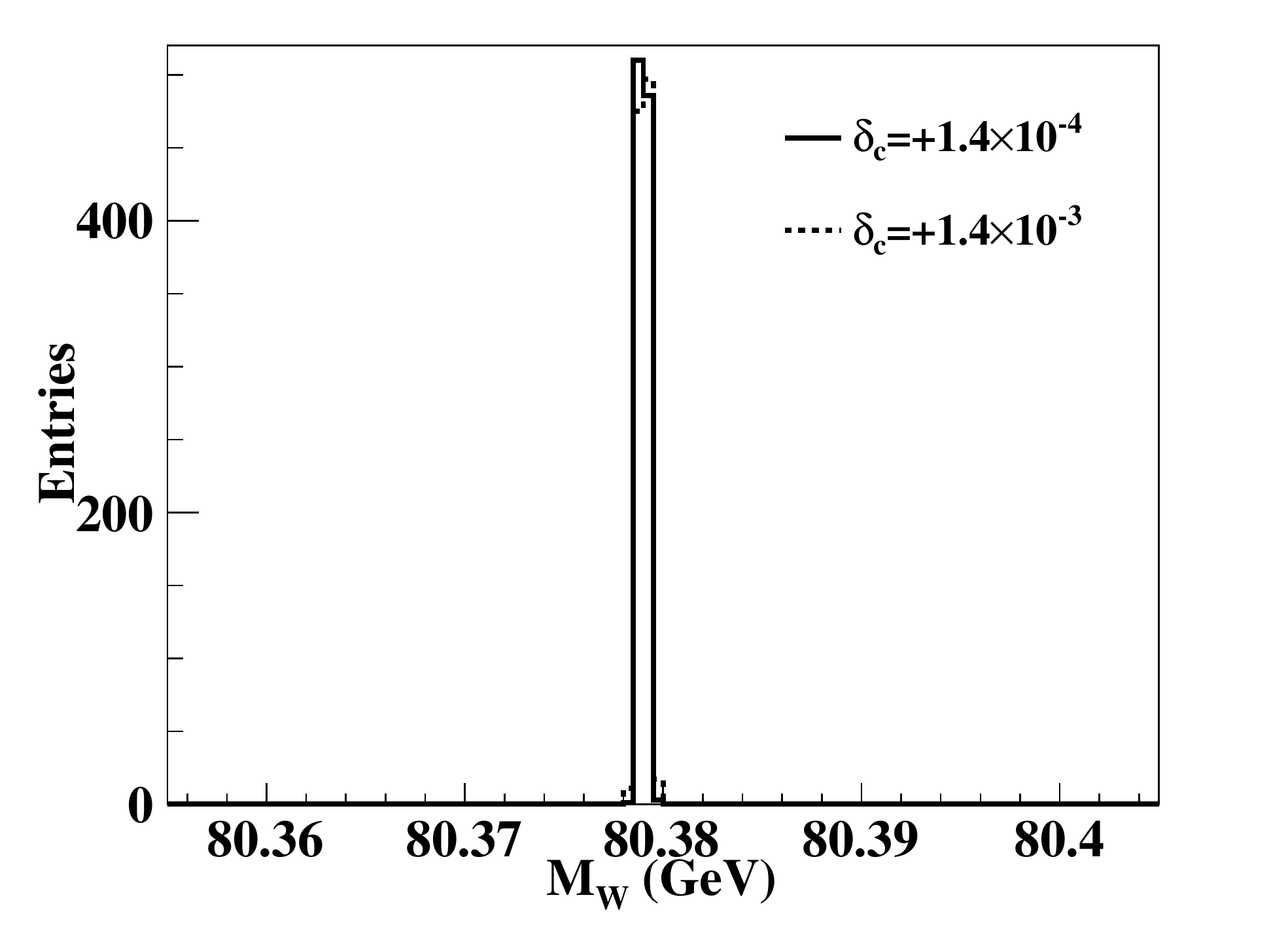}\put(-150,105){\bf (a)}
	}
	\subfigure{
		 \includegraphics[width=0.4\textwidth]{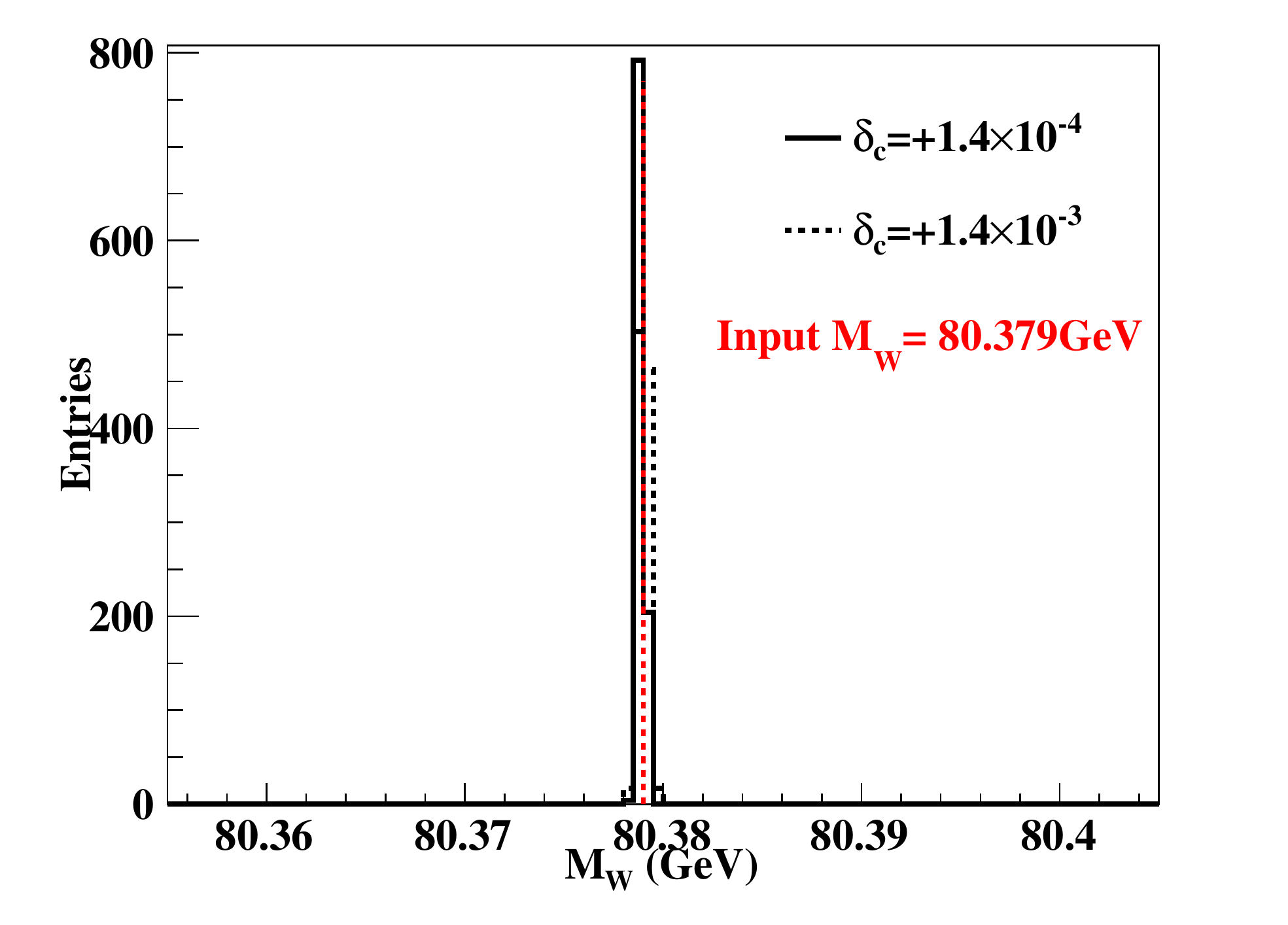}\put(-150,105){\bf (b)}
	}
	\caption{The simulation results of $m_{W}$ for the Gaussian (a) and non-Gaussian (b) cases,
        the input $m_{W}$ is 80.379~GeV and the scale factor method is used. The black solid and
        dash lines are the result for  $\delta_{c}=+1.4\cdot10^{-4}$ and $\delta_{c}=+1.4\cdot10^{-3}$
        at 161.2~GeV and 162.5~GeV, respectively.}
	\label{fig:SysCorr_chisq}
\end{figure}

The scale factor method is adopted for both the Gaussian and non-Gaussian cases.
Since an additional fit parameter, $h$, is needed for this method, the energy point at 162.5~GeV is added.
Figure~\ref{fig:SysCorr_chisq}(a) shows the simulation results for the Gaussian case.
We can see that even when $\delta_{c}$ increases by 10 times, the corresponding variation on $\Delta m_{W}$
is still very small. The advantage of this method is more obvious for the non-Gaussian case, as shown in
Fig.~\ref{fig:SysCorr_chisq}(b). Even though the uncertainties are increased by an order of magnitude,
the shift of $m_{W}$ is well controlled.

Apart from the uncertainties discussed above, the one on the theoretical calculated $W$-pair cross section
, $\Delta \sigma_{WW}$, is an important source which may limit the precision of the measured $m_{W}$ ($\Gamma_{W}$).
The contribution of $\Delta \sigma_{WW}$ is in the same form as Eq.~\ref{Eq_ErrLum}, and will have a prominent
effect if $\Delta \sigma_{WW}$ is quite large. With the $\Delta \sigma_{WW}$ at $10^{-3}$ level, the corresponding
$\Delta m_{W}$ is about 1.7~MeV at 161.2~GeV, which dilutes the statistical power obviously. The precision of
$\sigma_{WW}$ at threshold in LEP2 era was at $10^{-2}$ level, so it is extremely important to improve this for the precise
measurement of $m_{W}$ ($\Gamma_{W}$) in future.

%% ============================================== Data-taking scheme =============================================
\section{Data-taking strategies}
\label{Data taking}
In the above discussion, the main sources of the uncertainties of $m_{W}$
($\Delta\Gamma_{W}$ for data-taking at more than one point) are studied, including both the statistical and systematic ones.
Generally, $\Delta m_{W}$ ($\Delta\Gamma_{W}$) depends on the energy of the data point,
and the statistical part is also limited by the integrated luminosity.
The optimization of the data-taking strategy is to determine the
number of data-taking points, the energy of each data point, and the allocation
of the integrated luminosity for a fixed total integrated luminosity.
The FCC-ee has investigated data-taking at one and two energy points to measure
$m_{W}$ and $\Gamma_{W}$~\cite{Fcc_ee_WMass}.
% without taking into account systematic uncertainties.
When the systematic uncertainties are taken into account, especially for the correlated ones,
more energy points are beneficial for an optimal measurement.

MC experiment method is used to optimize the data-taking schemes.
%The production cross section of $W$-pair, the detection efficiency, the purity, and the luminosity,
%are used to get the number of $W$-pair events, as well as the uncertainties.
The number of $W$-pair events is compared with the theoretical predictions,
and the corresponding $m_{W}$ ($m_{W}$ and $\Gamma_{W}$) and its (their) uncertainties
can be obtained. The $\chi^{2}$ construction is listed in Eq.~\ref{Chi2_1} for data taking at one
or two energy points, and in Eq.~\ref{Chi2_Corr} for three energy points.

For each MC experiment, the statistical and uncorrelated systematic uncertainties,
are assumed to follow independent Poisson and Gaussian distributions at all energy points,
respectively; and for each correlated systematic uncertainty, the Gaussian distribution
is assumed. The experiments are repeated 500 times,
the corresponding distributions of $m_{W}$ and $\Gamma_{W}$ are expected to
follow Gaussian distribution, whose standard deviation represent
the combinations of all different uncertainty sources.

%% ============================================== Data taking at one point =============================================
\subsection{Measurement of the $W$ boson mass at one energy point}
\label{One_point}
\begin{figure}[htbp]
	\centering
	\subfigure{
		\includegraphics[width=0.45\textwidth]{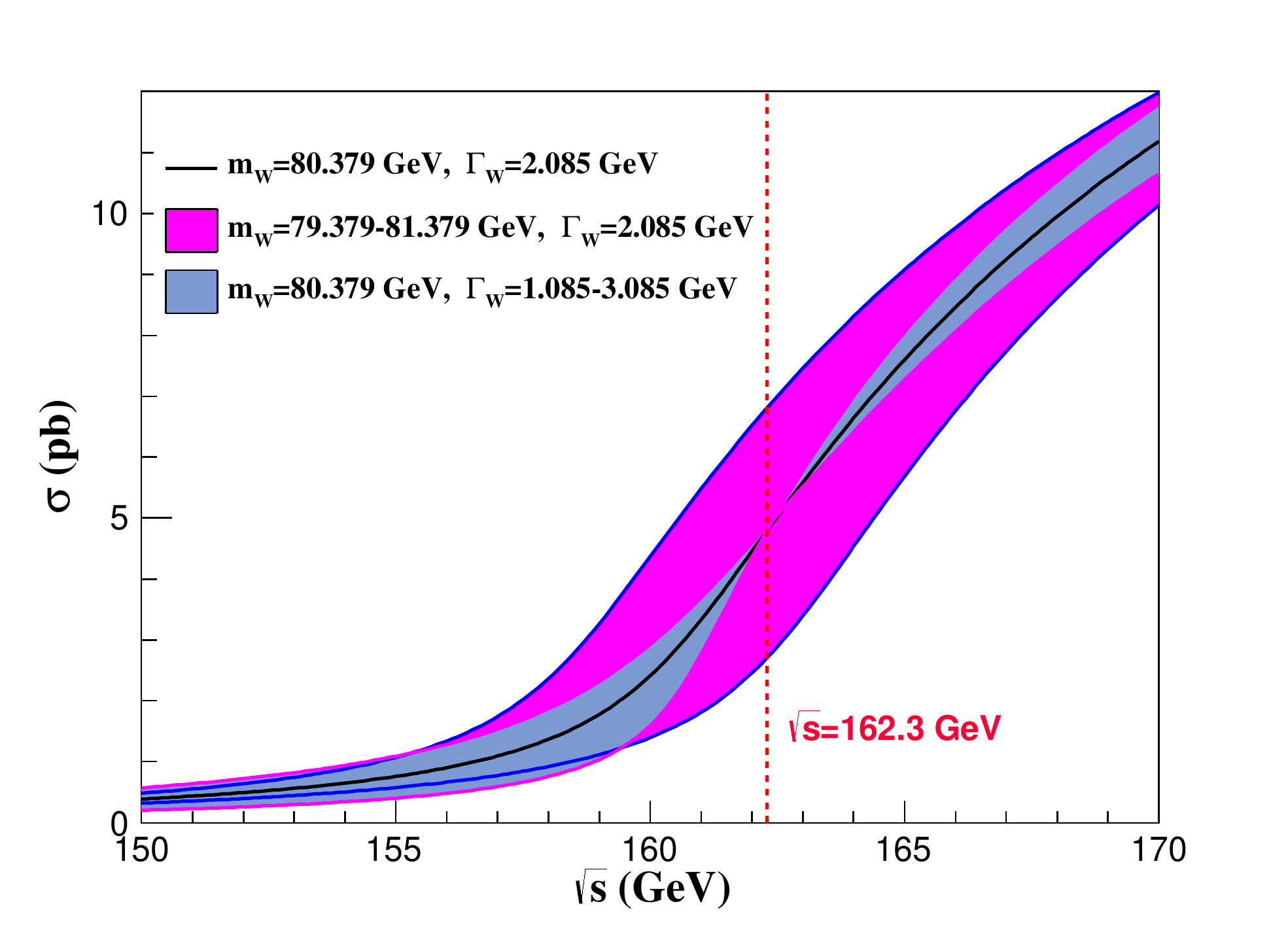}
	}
	\caption{The distribution of $W$-pair cross section as a function of $\sqrt{s}$.
			 The central curve corresponds to the result of using the PDG values of $m_{W}$ and
			 $\Gamma_{W}$~\cite{WMass_PDG2018}. Purple and green bands show the cross sections
             obtained by varying $m_{W}$ and $\Gamma_{W}$ within $\pm 1$~GeV.}
	\label{fig:Xsection_MW}
\end{figure}

For data taking at a single energy point,
there is an ideal choice, $E=2m_{W}+0.4\approx 161.2$~GeV, to measure $m_{W}$ with the best
statistical sensitivity as shown in the Fig.~\ref{fig:Stat_MW} (a).
But the contributions from systematic uncertainties need to be included for a realistic measurement.
%especially for the added source, $\Gamma_{W}$.
An interesting feature is the effect of the $\Gamma_W$ uncertainty on the $W$ boson mass.
Figure~\ref{fig:Xsection_MW} shows how the line-shape of $W$-pair cross section varies according to the $W$ boson mass and width,
where the black line is the one with $m_W$ and $\Gamma_W$ fixed to the world averaged values~\cite{WMass_PDG2018},
$m_{W}=80.385$~GeV and $\Gamma_{W}=2.085$~GeV,
and bands correspond to the variations of the $W$ boson mass or width in large ranges, $\pm 1$~GeV.
It can be seen that although the variation of the $W$ boson width changes the cross section line-shape,
there is a common intersection of all the line-shape curves with different $\Gamma_{W}$, $\sqrt{s} = 162.3$~GeV,
which indicates that the cross section around this energy points is insensitive to the uncertainty of the $W$ boson width.

Based on the above discussion,
two specific energy points are favored for the $W$ mass measurement.
The first is the most statistically sensitive one, $E=161.2$~GeV,
and the other is $E=162.3$~GeV,
where the uncertainties of $\Gamma_{W}$ and the $E_{BS}$ have negligible effects on the $W$  mass measurement.
At 161.2~GeV, the effect of uncertainty from $\Gamma_{W}$ on $W$ mass can be written as
$\Delta m_{W} = (\frac{\partial m_{W}}{\partial \sigma_{WW}}) \Delta \sigma_{WW}
 = (\frac{\partial m_{W}}{\partial \sigma_{WW}}) (\frac{\partial \sigma_{WW}}{\partial \Gamma_{WW}}) \Delta \Gamma_{W}$,
which could be estimated with numerical calculation. With $\frac{\partial m_{W}}{\partial \sigma_{WW}}= 0.474 \rm{GeV/pb}$
and $\frac{\partial \sigma_{WW}}{\partial \Gamma_{WW}}=0.376\rm{pb/GeV}$ at 161.2GeV and $\Delta \Gamma_{W}=42$~MeV~\cite{WMass_PDG2018},
the corresponding $\Delta m_{W}$ is about 7.5~MeV.

Table~\ref{tab:DT_OneP} summarizes the results for the data taking at either one
of the above two energy points with the configurations in Table~\ref{tab:configurations}.
It can be seen that the dominant contribution to $\Delta m_{W}$ at 161.2~GeV is from the uncertainties of $\Gamma_{W}$,
which is negligible at $E=162.3$~GeV.
Thus 162.3~GeV is a better choice when only $m_{W}$ is measured and the expected precision is about $0.9\oplus\rm{theory~MeV}$.

\begin{table*}[htbp]
	\begin{center}
		\caption{The precision of $m_{W}$ when taking data at $E=161.2$ or $162.3$ GeV.
		Shown in the table are the statistical and systematic uncertainties on $m_W$. The last column is the total uncertainty at the
		corresponding energy point.}
		\begin{tabular}{c|c|c|ccccc|c}
			\hline
			\multicolumn{2}{c|}{Energy/source} & $\delta_{\rm{stat}}$ (stat.) & $\Delta E$ & $\Delta \sigma_{E}$
			& $\Delta\Gamma_{W}$ & $\delta_{\rm{B}}$ & $\delta_{c}$ & Total \\
			\hline
			\multirow{2}{2cm}{$\Delta m_{W}$ (MeV)} & 161.2 (GeV) & 0.59 & 0.36 & 0.20 & 7.49 & 0.17 & 0.24 & 7.53\\
            \cline{2-9}
			& 162.3 (GeV) & 0.65 & 0.37 & -    & -   & 0.17 & 0.34 & 0.84\\
			\hline
		\end{tabular}
		\label{tab:DT_OneP}
	\end{center}
\end{table*}

%% ============================================== Data-taking at two points =============================================
\subsection{Measurement of the $W$ boson mass and width at two energy points}
\label{Two_points}
In the previous section, data taking at one energy point is
investigated, the best precision of $m_{W}$ can be obtained with $E=162.3$~GeV.
With one energy point, only $m_{W}$ can be measured.
Alternately, both $m_{W}$ and $\Gamma_{W}$ can be determined simultaneously
if two energy points near the $W$-pair threshold are adopted for data-taking.
In this case, the statistical uncertainties of $m_{W}$ and $\Gamma_{W}$
can be obtained using Eq.~\ref{EMatrix}.

To obtain the best precision of $m_{W}$ and $\Gamma_{W}$ for a given total integrated luminosity,
the data-taking scheme of the energy points
and luminosity allocation for each energy point are optimized.
A 3-dimensional (3D) scan of the energy points $E_{1}$ and $E_{2}$ ($E_{1}<E_{2}$),
and the luminosity fraction $F$ of the energy point $E_{1}$ is performed,
which defined as $F=\mathcal{L}_{1}/\mathcal{L}$. The scan step sizes of
$E_{1}$ and $E_{2}$ are 100~MeV, and 0.05 for $F$.

\begin{figure}[htbp]
	\centering
	\subfigure{
		 \includegraphics[width=0.3\textwidth]{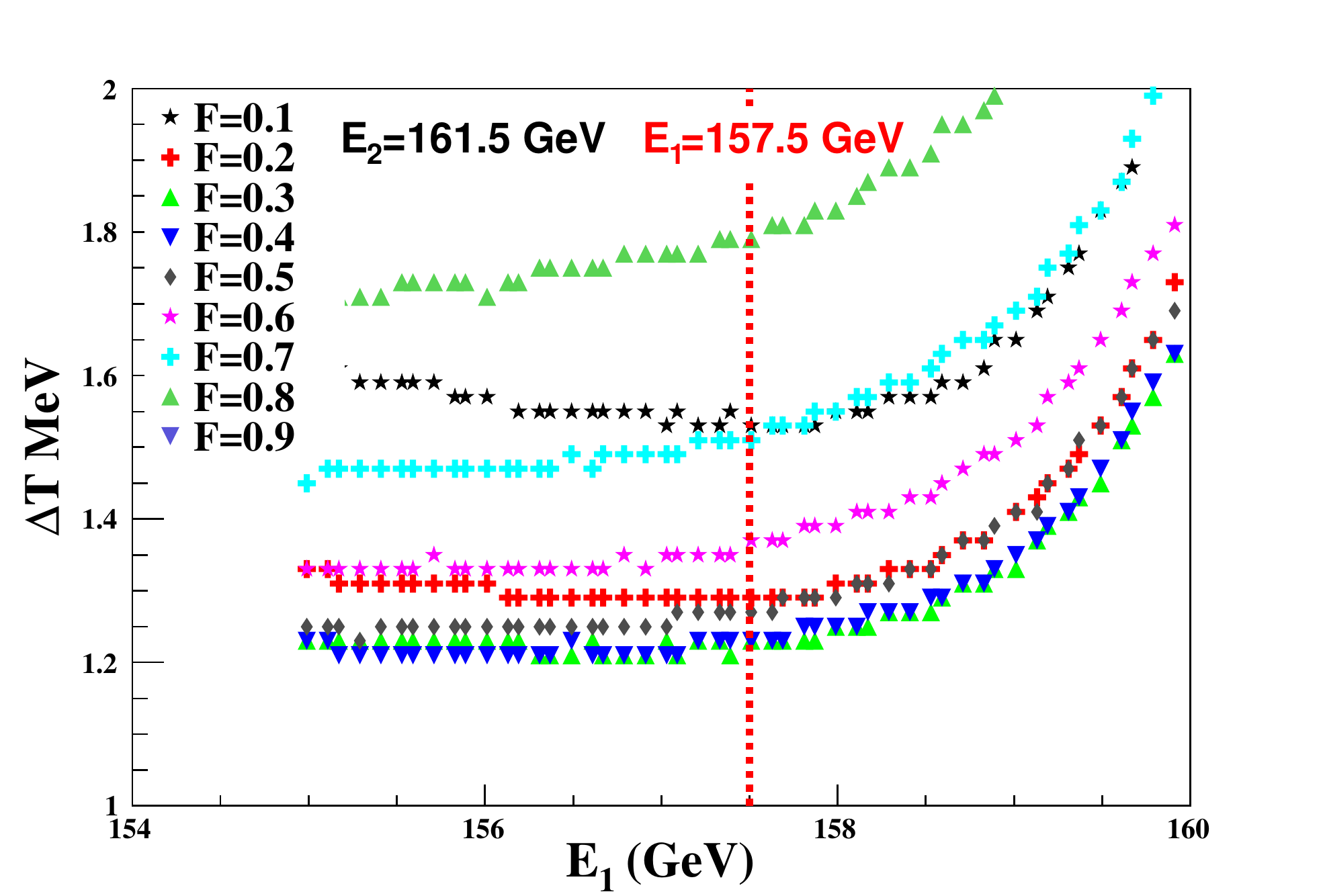}\put(-160,40){\bf (a)}
	}
	\subfigure{
		 \includegraphics[width=0.3\textwidth]{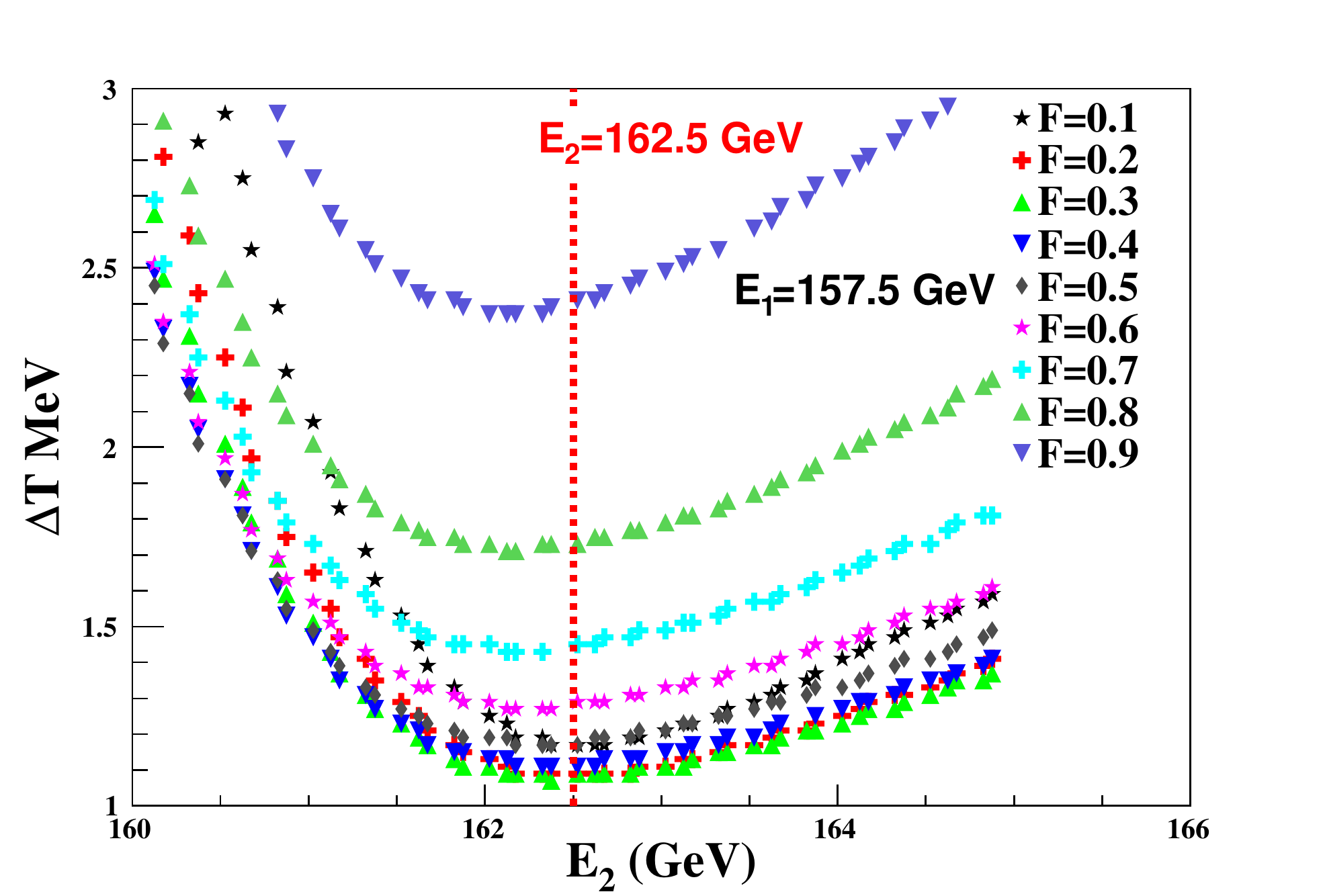}\put(-160,40){\bf (b)}
	}
	\subfigure{
		 \includegraphics[width=0.3\textwidth]{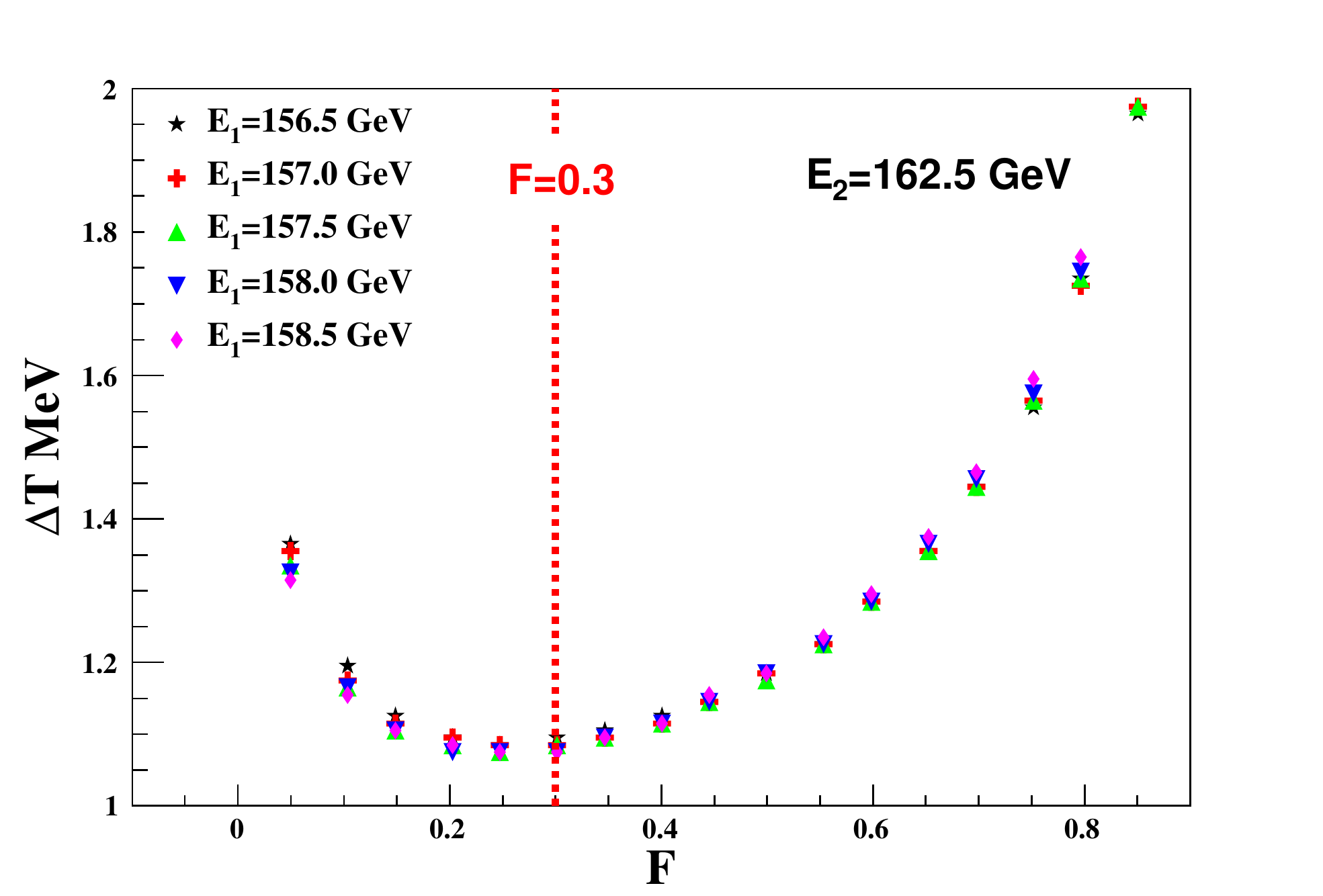}\put(-160,40){\bf (c)}
	}
	\caption{The optimization results of 3D scan for taking data at two points.
	(a)-(c) are for $E_{1}$, $E_{2}$, and $F$, respectively.
    In practice, each parameter is optimized by scanning other two parameters.
    These three plots just shows the dependence of $\Delta T$ on one parameter, with
    another one float and the third one fixed.}
	\label{fig:DT_TwoP}
\end{figure}

The best energy point for $m_{W}$ is above the $W$-pair threshold,
while the one for $\Gamma_{W}$ is below the threshold, as shown in Fig.~\ref{fig:Stat_MW}, making it
impossible to simultaneously achieve the best precisions for both.
Thus an objective function is defined to quantify the relative importance of the two measurements:
$T=m_{W} + A\cdot\Gamma_{W}$, where $A$ is the weight factor to be chosen.
Since the $W$ boson mass is thought to be more important than its width, $A=0.1$ is used
throughout this paper, and the goal of optimization is to minimize $\Delta T$.
%During the optimization of the scan parameter such as $E_{1}$, the different dependences of $\Delta T$ on $E_{1}$ are obtained by scanning other two parameters, than the value with minimum $\Delta T$ is taken  as the optimized result for $E_{1}$.
Figure~\ref{fig:DT_TwoP} (a)-(c) show the optimization of $E_{1}$, $E_{2}$, and $F$.
%In the optimization, a 3D scan is performed, and these plots are used to illustrate the final results.
%When the dependence of $\Delta T$ on one parameter is plotted, another one is fixed with the scanning of the third one.
For the scheme of two energy points, the optimized parameter values are:
\begin{equation}
E_{1} = 157.5~\rm{GeV},~~~E_{2} = 162.5~\rm{GeV},~~~F = 0.3,
\end{equation}
where $E_2=162.5$ GeV is consistent with the expectation,
since $\Delta m_{W}$ is minimal around this energy region
and has more weight to $\Delta T$. The $W$-pair cross section is not very sensitive to $m_{W}$ when $\sqrt{s}$ is
less than 158~GeV, thus the distribution of $\Delta T$ is generally flat in this energy
region. The corresponding luminosity fraction is smaller than the one around $162.5$~GeV.
%With these results, together with the configurations of total integrated luminosity and the  control of the systematic uncertainties,
The projected precisions for $m_{W}$ and $\Gamma_{W}$ are summarized in Table~\ref{tab:DT_TwoP}.

\begin{table*}[htbp]
		\caption{The expected precisions of $m_{W}$ and $\Gamma_{W}$ with the optimized data-taking schemes.
		Listed are the effects of different uncertainty  sources such as statistical,
		un-correlated systematic ($\Delta E$ and $\Delta\sigma_{E}$), and correlated systematic.
		The last column shows the total uncertainties on the $W$ mass and width. }
		\begin{center}
		\begin{tabular}{c|c|c|cccc|c}
			\hline
			\multirow{2}{2.0cm}{Data-taking scheme} & \multirow{2}{2.0cm}{mass or width} & \multirow{2}{1.7cm}{$\delta_{\rm{stat}}$~(MeV)}
            & \multicolumn{4}{c|}{$\delta_{\rm{sys}}$~(MeV)} & \multirow{2}{1.7cm}{Total~(MeV)} \\
            \cline{4-7}
            & & & $\Delta E$ & $\Delta \sigma_{E}$ & $\delta_{B}$ & $\delta_{c}$ & \\
			\hline
			One point & $\Delta m_{W}$& 0.65 & 0.37 & - & 0.17 & 0.34 & 0.84 \\
			\hline
			\multirow{2}{1.7cm}{Two points} & $\Delta m_{W}$   & 0.80 & 0.38 & - & 0.21 & 0.33 & 0.97\\
										  & $\Delta\Gamma_{W}$ & 2.92 & 0.54 & 0.56 & 1.38 & 0.20 & 3.32\\
			\hline
			\multirow{2}{1.95cm}{Three points} & $\Delta m_{W}$ & 0.81 & 0.30 & - & 0.23& 0.29 & 0.98\\
										  & $\Delta\Gamma_{W}$  & 2.93 & 0.52 & 0.55 & 1.38 & 0.20 & 3.37\\
			\hline
		\end{tabular}
		\label{tab:DT_TwoP}
	\end{center}
\end{table*}

%% ============================================== Data taking at three points =============================================
\subsection{Measurement of the $W$ boson mass and width at three energy points}
\label{Three_points}

For taking data at more than two energy points near the $W$-pair threshold,
the correlation in the $m_{W}$ and $\Gamma_{W}$ measurements among different energy points
can be taken into account by redefining the $\chi^{2}$ form and introducing additional parameter(s)
$h_i$ as shown in Eq.~\ref{Chi2_Corr}.
Therefore the effects of the correlated systematic uncertainties are reduced, leading to improved
precisions of the measurements.

The procedure of optimization for three energy points scheme is
analogous to the case for two energy points by adding another two scan parameters.
The energies of the three data points, $E_{1}$, $E_{2}$, and $E_{3}$,
as well as the two luminosity fractions $F_{1}$ and $F_{2}$ are optimized
to reach the best precisions of $m_{W}$ and $\Gamma_{W}$,
where $F_{1}=\mathcal{L}_{1}/\mathcal{L}$ and $F_{2}=\mathcal{L}_{2}/\mathcal{L}$.
The scan procedure is similar to that for the two energy points, except it is over a 5-dimensional
parameter space now. The optimized parameter values are:
\begin{equation}
\begin{aligned}
	\label{Optimize_3P}
	&E_{1} = 157.5~\rm{GeV},~E_{2} = 162.5~\rm{GeV}, \\
    &E_{3}=161.5~\rm{GeV},~F_{1} = 0.30,~F_{2}=0.63.
\end{aligned}
\end{equation}

With these results and the assumptions of total integrated luminosity and the systematic uncertainties,
the expected $\Delta m_{W}$ and $\Delta \Gamma_{W}$  are listed in Table~\ref{tab:DT_TwoP},
and the total projected uncertainties would be
\begin{equation}
    \Delta m_{W} \sim 1.0\oplus \Delta_{M}^{th}\rm{~MeV},~~~\Delta\Gamma_{W}\sim 3.4\oplus \Delta_{\Gamma}^{th}\rm{~MeV},
\end{equation}
where$\Delta_{M}^{th}$ and $\Delta_{\Gamma}^{th}$ are the theoretical uncertainties of the $W$ boson mass and width due to the cross section calculation
Though the precisions of the $W$ boson mass and width for the three energy points are not improved
much compared with those for the two energy points, the results for the three energy points are
more realistic and robust. Since more energy points have the advantage of better background understanding
and the sophisticated treatment of correlated systematic uncertainties.

%% ============================================== Discussion  =============================================
\subsection{Discussion about the data-taking plan}

Three data-taking schemes are investigated above for the best measurement precisions of the
$W$ boson mass and width with the threshold scan method. With the fixed total integrated
luminosity and expectations on systematic uncertainty controls, the data-taking is optimized
to minimize the total uncertainties on the $W$ boson mass and width measurements.
%, which is to balance the statistical and systematic uncertainties.

The integrated luminosities of the CEPC and the FCC-ee at the $W$-pair threshold are expected to
be much larger than that at the LEP. In the ideal case of one single energy point,
both the analytic and MC simulation method have showed that a statistical precision of less than
1~MeV can be achieved for $m_W$. It indicates that the systematic uncertainties
such as theoretical calculation, beam energy calibration, luminosity determination, {\it etc.} become more important.
One interesting feature is that the $\Delta m_W$ due to the $W$ boson width and the beam
energy spread vanishes around $\sqrt{s} = 2m_W+1.5$~GeV. These two systematic uncertainties can be neglected for the data taking at this energy point.

For taking data at a single energy point, the $W$ boson mass and width cannot be determined simultaneously. Moreover, the best precision of either is obtained at different energies.
However, the optimized $\Delta m_{W}$ for the two or three energy points is
only slightly larger than the one for a single energy point as shown in Table~\ref{tab:DT_TwoP}.
In this case,  $\Gamma_{W}$ can be measured simultaneously.
Also, although the optimized precisions on $m_{W}$ and $\Gamma_{W}$ are similar for the two and three energy points,
the latter is beneficial for the treatment of the correlated systematic uncertainties,
especially when the effects of these uncertainties are in the absolute form, which will cause shifts to the obtained
$m_{W}$ and $\Gamma_{W}$. Therefore, data taking at three different energy points is preferred, the corresponding optimal
data-taking scheme is listed in  Eq.~\ref{Optimize_3P}.

Thanks to contributions from FCC-ee~\cite{EPOL17,Fcc_ee_WMass} studies,
the different types of systematic uncertainties are considered comprehensively in this work,
and the numerical results of the contributions of the dominant backgrounds are estimated.
In this paper, the data taking schemes are optimized for a total integrated luminosity of 3.2~$\mbox{ab}^{-1}$~\cite{CEPC_1}.
The results of the optimization can be scaled to other integrated luminosities. Table~\ref{tab:Scale_L} lists the precisions of $m_{W}$ and $\Gamma_{W}$ with the threshold scan method, varying the total luminosity between 1~$\mbox{ab}^{-1}$ and 15$~\mbox{ab}^{-1}$. The three data taking schemes in the table are the optimized
results described above, and all the uncertainties are statistical only. One can obtain the total uncertainty by adding the systematic uncertainties summarized in Table~\ref{tab:DT_TwoP}.
The results for an integrated luminosity of 15~$\mbox{ab}^{-1}$ are comparable with FCC-ee's results:
(1) for the one energy point scheme,
our result of $\Delta m_W = 0.31$~MeV at 162.3~GeV is slightly worse than that of the FCC-ee study, {\em i.e.}, 0.25~MeV at 161.4~GeV .
Since the uncertainty of $\Gamma_{W}$  has significant contribution to $\Delta m_W$ around the most statistically sensitive energy point (up to 8~MeV),
so the one at 162.3~GeV is chosen in this work, where the $W$-pair cross section is insensitive to the $\Gamma_{W}$
and the statistical uncertainty of the $m_{W}$ increases a bit.
(2) for the two energy points scheme, since the $W$ mass is thought to be more important than its width,
it's reasonable to allocate more luminosity to the energy point that benefits the $m_{W}$ measurement.
So the precision of $m_{W}$ is slightly better than FCC-ee's result, contrary to the precision on $\Gamma_{W}$.
It is worth noting that the contribution to  $\Delta m_{W}$ from systematic uncertainties will
become more important with the increasing of the luminosity, so the consideration of the systematic uncertainties
is more important. With this in mind, the three energy points data-taking scheme is preferred since it
allows for better control and treatment of the systematic uncertainties.

\begin{table*}[htbp]
		\caption{The expected precisions of $m_{W}$ and $\Gamma_{W}$ with the optimized data-taking schemes,
        corresponding to different luminosity inputs (statistical uncertainties only). The last column is the result of
        FCC-ee, where the systematic uncertainties are reckoned to be under control to a negligible level of impact~\cite{Fcc_ee_WMass}.
        Several systematic uncertainties have been studied in section~\ref{Section_sys}, and the numerical results with
        the assumptions listed in Table~\ref{tab:configurations} can be found in Table~\ref{tab:DT_OneP} and ~\ref{tab:DT_TwoP}.
        It should be noted that the uncertainty related to the theoretical precision of the $W$-pair cross section is not include.
        Its improvement is necessary for the high precision measurement of $m_{W}$ ($\Gamma_{W}$), as well as the controlling
        of other systematic uncertainties.
        }
		\begin{center}
		\begin{tabular}{c|c|ccccc|c|c}
			\hline
			\multirow{2}{1.8cm}{Data-taking scheme} & \multirow{2}{2.0cm} {mass or width~(MeV)} & \multicolumn{6}{c|}{Luminosity (ab$^{-1}$)} & FCC-ee\\
            \cline{3-9}
             & & 1 & 3 & 6 & 9 & 12 & 15 & 15\\
			\hline
			One point & $\Delta m_{W}$& 1.15 & 0.67 & 0.47 & 0.39 & 0.33 & 0.30 & 0.25\\
			\hline
			\multirow{2}{1.7cm}{Two points} & $\Delta m_{W}$&    1.42 & 0.82 & 0.58 & 0.47 & 0.41 & 0.37 & 0.41\\
										  & $\Delta\Gamma_{W}$& 5.21 & 3.02 & 2.13 & 1.74 & 1.51 & 1.35 & 1.10\\
			\hline
			\multirow{2}{1.95cm}{Three points} & $\Delta m_{W}$& 1.43 & 0.82 & 0.58 & 0.48 & 0.41 & 0.37 & -\\
										  & $\Delta\Gamma_{W}$& 5.24 & 3.02 & 2.14 & 1.75 & 1.51 & 1.35  & -\\
			\hline
		\end{tabular}
		\label{tab:Scale_L}
	\end{center}
\end{table*}

%% ============================================== Summary =============================================
\section{Summary}
\label{Summary}
In this paper, different data-taking schemes are investigated for the precise measurements of the $W$ boson mass and width
at further circular electron positron colliders, such as the CEPC and FCC-ee.
For a fixed total integrated luminosity, $\mathcal{L}=3.2~\mbox{ab}^{-1}$,
and the expectations of the systematic uncertainties, taking data at three energy points is found to be optimal
with the energies and luminosity allocations listed in Eq.~\ref{Optimize_3P}.
The corresponding projected uncertainties on the $W$ boson mass and width are
$\Delta m_{W}\sim 1.0$~MeV and $\Delta\Gamma_{W}\sim 3.4$~MeV, respectively.
Various systematic uncertainties are taken into account in the investigation.
The one related to the theoretical calculation of the $W$-pair cross section is discussed but
not included in the numerical results and listed separately. It's critical to improve the calculation for the high precision measurement
of $m_{W}$ ($\Gamma_{W}$) using the threshold scan method.

%%%%%%%%%%%%%%%%%%%%%%%%%%%%%%%%%%%%%%%%%%%%%%%%%%%%%%%%
%===========================================================================
%======================== acknowledgments ==================================
%===========================================================================
\begin{acknowledgements}
This work is supported in part by National Key Basic Research Program of China (Grant No. 2017YFA042203);
National Natural Science Foundation of China (NSFC) (Grant No. 11875278);
National Key Program for S\&T Research and Development of China (Grant No. 2016YFA0400400);
Joint Large-Scale Scientific Facility Funds of the NSFC and CAS (Grant Nos. U1332201, U1532258);
the Hundred Talents Program of the Chinese Academy of Sciences (CAS) (Grant No. Y6291150K2);
Beijing Municipal Science and Technology Commission project (Grant Nos. Z191100007219010, Z181100004218003).
\end{acknowledgements}

% ==================================================  References  ======================================================

\end{document}